\documentclass[fleqn,usenatbib]{mnras}

\usepackage[T1]{fontenc}
\usepackage{ae,aecompl}
\usepackage{natbib}

\def\fesc{\ifmmode f_{\rm esc} \else $f_{\rm esc}$\fi}

\usepackage{graphicx}	

\title[Ly$\alpha$ emission in XMDs]{Ly$\alpha$ emission in low-redshift most metal-deficient compact star-forming galaxies}

\author[Y. I. Izotov et al.]{
Y. I. Izotov$^{1}$, 
T. X. Thuan$^{2}$,
N. G. Guseva$^{1}$, 
D. Schaerer$^{3,4}$, 
G. Worseck$^{5}$, 
\newauthor 
~A. Verhamme$^{3}$
\\
$^{1}$Bogolyubov Institute for Theoretical Physics,
National Academy of Sciences of Ukraine, 14-b Metrolohichna str., Kyiv,
03143, Ukraine,\\
E-mail: yizotov@bitp.kiev.ua, nguseva@bitp.kiev.ua\\
$^{2}$Astronomy Department, University of Virginia, P.O. Box 400325, 
Charlottesville, VA 22904-4325, USA,\\
E-mail: txt@virginia.edu\\
$^{3}$Observatoire de Gen\`eve, Universit\'e de Gen\`eve, 
51 Ch. des Maillettes, 1290, Versoix, Switzerland,\\
E-mail: daniel.schaerer@unige.ch, anne.verhamme@unige.ch\\
$^{4}$IRAP/CNRS, 14, Av. E. Belin, 31400 Toulouse, France\\
$^{5}$ Institut f\"ur Physik und Astronomie, Universit\"at Potsdam, Karl-Liebknecht-Str. 24/25, D-14476 Potsdam, Germany,\\
E-mail: gworseck@web.de
}

\date{Accepted XXX. Received YYY; in original form ZZZ}

\pubyear{2019}

\begin{document}
\label{firstpage}
\pagerange{\pageref{firstpage}--\pageref{lastpage}}
\maketitle

\begin{abstract}
We present observations with the Cosmic Origins Spectrograph onboard the {\sl Hubble Space Telescope} of nine most metal-deficient compact star-forming galaxies with oxygen abundances 12 + log(O/H) = 6.97 --  7.23, redshifts $z$ = 0.02811 -- 0.13320, and stellar masses $M_\star$ $\leq$ 10$^7$ M$_\odot$. We aim to study the properties of Ly$\alpha$ emission in these extremely metal-deficient objects. We find that all nine galaxies are Ly$\alpha$ emitters (LAEs). We examine various relations between the Ly$\alpha$ escape fraction $f_{\rm esc}$(Ly$\alpha$) and other characteristics -- such as absolute UV magnitude, oxygen abundance, O$_{32}$ ratio, stellar mass, Lyman-alpha luminosity and equivalent width EW(Ly$\alpha$), ionizing photon production efficiency $\xi_{\rm ion}$ and velocity separation $V_{\rm sep}$ between the two peaks of the Ly$\alpha$ profile  -- of a large sample of LAEs, including our lowest-metallicity galaxies and other objects from the literature. We find a relatively tight correlation between $f_{\rm esc}$(Ly$\alpha$) and two characteristics, EW(Ly$\alpha$) and $V_{\rm sep}$, whereas no correlation is found between $f_{\rm esc}$(Ly$\alpha$) and the oxygen abundance. We also find a relatively tight relation between the Ly$\alpha$ and LyC escape fractions. We propose to use the latter relation to estimate indirectly the escaping ionizing radiation in LAEs, when direct measurements of LyC emission are not possible. We show that the global properties of low-$z$ LAEs are very similar to those of $z$ $>$ 6 galaxies. They are thus ideal local proxies for studying physical processes during the epoch of reionization of the Universe.
\end{abstract}




\section{Introduction}\label{intro}

Theoretical and observational studies over the last decade have
suggested that the agents responsible for 
the reionization of the Universe, which ends at $z$~$\sim$~6, are likely to come from 
a dominant population of low-mass and very metal-poor dwarf galaxies
\citep*{O09,WC09,K11,J13,M13,Y11,B15a,Sm18,N18,St18,K19}. 
On the observational side, 
recent work including some of the first observations of the
{\sl James Webb Space Telescope} ({\sl JWST}) revealed large populations
of UV-faint galaxies at $z$ $>$ 6 with low stellar masses
($M_\star$~$\sim$~10$^6$~--~10$^8$~M$_\odot$), active star formation with specific
star-formation rates sSFR of $\sim$ 100 Gyr$^{-1}$, and likely extremely
low metallicities, although the latter are to be confirmed by further observations 
\citep[e.g. ][]{Pe18,Fu20,Ki20,Sc22,Cu23,Ni23,En23,En23b,Sa23,Ma23,M23,Si23,San23,Tr23,Ch23,Rh23,Fu23,Lin23,A23}.
These galaxies are characterized by high ionizing
photon production efficiencies $\xi_{\rm ion}$ of
$\sim$ 10$^{25.5}$~--~10$^{25.8}$~Hz~erg$^{-1}$, 
and are likely numerous enough to be able to reionize the Universe. Many of these galaxies show 
strong Lyman-alpha emission, and so are classified as  
Lyman-alpha emitters (LAEs)
\citep[e.g. ][]{Jo23,Sa23,Ta23,Ju23,Ni23}.

However, despite these recent advances, those dwarf
star-forming galaxies (SFGs) at high redshfifts are very faint and most of
them still escape detection or do not allow detailed studies by modern large telescopes.
Therefore, it is important to search for local counterparts of 
these high-redshift dwarf SFGs to study their properties in detail.

In particular, the profiles of the resolved Ly$\alpha$ emission line are
used as indirect indicators of escaping LyC emission. These profiles in
the spectra of most of low-$z$ LyC leakers and LAEs are double-peaked,
with the peak separation $V_{\rm sep}$ increasing with increasing
optical depth in the Ly$\alpha$ emission line, which depends on the neutral
hydrogen column density $N$(H~{\sc i}) \citep[e.g. ][]{I16,I16b,I18,I18b}. 
Only in a few galaxies, the Ly$\alpha$ profile consists of
more peaks \citep{RT17,I18b,Va18}.

The relation between  the LyC escape fraction $f_{\rm esc}$(LyC) and
the Ly$\alpha$ peak separation is tight \citep{V15,I18b}, making $V_{\rm sep}$
the reliable indicator of escaping LyC emission.
Furthermore, \citet{V17} have shown that the Ly$\alpha$ escape fractions
$f_{\rm esc}$(Ly$\alpha$) are larger than the LyC escape fractions 
$f_{\rm esc}$(LyC), in accord with theoretical predictions of \citet*{D16}
and simulations of \citet{Ma22}.
A more detailed analysis of simulated Lyman-alpha profiles is given in
\citet{Bl23}.
\citet{I20} found that the Ly$\alpha$ escape fraction also anticorrelates
with $V_{\rm sep}$, but not as tightly as $f_{\rm esc}$(LyC).

  \begin{table*}
  \caption{Some general characteristics of the selected galaxies from the SDSS data base
\label{tab1}}
\begin{tabular}{lrrccccccc} \hline
Name&R.A.(2000.0)&Dec.(2000.0)&$z$&$D_L$$^{\rm a}$&$D_A$$^{\rm b}$&O$_{32}$$^{\rm c}$
&EW(H$\beta$)&FWHM$^{\rm d}$&12+log(O/H)$^{\rm e}$ \\ 
    &            &            &   & (Mpc)       &  (Mpc)      &
&(\AA)&(arcsec)& \\ \hline
J0122$+$0048&01:22:41.62&$+$00:48:42.06&0.05734&258& 231& 4.8&156&1.52&7.22\\ 
J0139$+$1542&01:39:11.93&$+$15:42:41.32&0.02811&124& 117& 5.9&339&1.19&7.22\\ 
J0811$+$4730&08:11:52.12&$+$47:30:26.24&0.04442&198& 181& 9.6&331&1.21&6.97\\ 
J0837$+$1921&08:37:21.87&$+$19:21:10.63&0.06734&305& 268&10.7&134&1.03&7.21\\ 
J1004$+$3256&10:04:09.90&$+$32:56:12.51&0.06639&301& 264&37.0&459&1.61&7.16\\ 
J1206$+$5007&12:06:08.53&$+$50:07:21.17&0.05129&230& 208& 5.6&219&1.67&7.15\\ 
J1234$+$3901&12:34:15.70&$+$39:01:16.41&0.13320&630& 491&14.1&276&1.28&7.03\\ 
J1505$+$3721&15:05:08.58&$+$37:21:40.22&0.07532&343& 297&15.8&298&1.40&7.23\\ 
J2229$+$2725&22:29:33.19&$+$27:25:25.60&0.07622&347& 300&54.8&580&0.93&7.11\\ 
\hline
\end{tabular}

\hbox{$^{\rm a}$Luminosity distance \citep[NED, ][]{W06}.}

\hbox{$^{\rm b}$Angular size distance \citep[NED, ][]{W06}.}

\hbox{$^{\rm c}$O$_{32}$ = [O~{\sc iii}]$\lambda$5007/[O~{\sc ii}]$\lambda$3727.}

\hbox{$^{\rm d}$Full width at half maximum in the SDSS $g$-band image.}

\hbox{$^{\rm e}$Oxygen abundance derived by the $T_{\rm e}$-method from line intensities in the SDSS spectrum.}

  \end{table*}

  \begin{table*}
  \caption{Apparent magnitudes with errors in parentheses compiled
from the SDSS, {\sl GALEX} and {\sl WISE} data bases
\label{tab2}}
\begin{tabular}{lccccccccccccc} \hline
Name&\multicolumn{5}{c}{SDSS (AB)}
&&\multicolumn{2}{c}{{\sl GALEX} (AB)}&&\multicolumn{4}{c}{{\sl WISE} (Vega)} \\ 
    &\multicolumn{1}{c}{$u$}&\multicolumn{1}{c}{$g$}&\multicolumn{1}{c}{$r$}&\multicolumn{1}{c}{$i$}&\multicolumn{1}{c}{$z$}&&FUV&NUV&&\multicolumn{1}{c}{$W1$}&\multicolumn{1}{c}{$W2$}&\multicolumn{1}{c}{$W3$}&\multicolumn{1}{c}{$W4$}
\\
    &\multicolumn{1}{c}{(err)}&\multicolumn{1}{c}{(err)}&\multicolumn{1}{c}{(err)}&\multicolumn{1}{c}{(err)}&\multicolumn{1}{c}{(err)}&&(err)&(err)&&\multicolumn{1}{c}{(err)}&\multicolumn{1}{c}{(err)}&\multicolumn{1}{c}{(err)}&\multicolumn{1}{c}{(err)} \\
\hline
J0122$+$0048& 22.02& 21.62& 22.18& 21.75& 22.19&& 21.90& 21.52&&\multicolumn{1}{c}{...   }&\multicolumn{1}{c}{...   }&\multicolumn{1}{c}{...   }&\multicolumn{1}{c}{ ... } \\
            &(0.17)&(0.06)&(0.13)&(0.14)&(0.60)&&(0.34)&(0.27)&&\multicolumn{1}{c}{(...) }&\multicolumn{1}{c}{(...) }&\multicolumn{1}{c}{(...) }&\multicolumn{1}{c}{(...)} \\
J0139$+$1542& 22.24& 21.84& 22.12& 22.39& 21.89&& 23.88& 22.17&&\multicolumn{1}{c}{ ...  }&\multicolumn{1}{c}{ ...  }&\multicolumn{1}{c}{ ...  }&\multicolumn{1}{c}{ ... } \\
            &(0.29)&(0.07)&(0.13)&(0.25)&(0.57)&&(0.28)&(0.15)&&\multicolumn{1}{c}{(...) }&\multicolumn{1}{c}{(...) }&\multicolumn{1}{c}{(...) }&\multicolumn{1}{c}{(...)} \\
J0811$+$4730& 22.30& 21.38& 22.17& 21.56& 23.05&& ...  & 22.79&&\multicolumn{1}{c}{ ...  }&\multicolumn{1}{c}{ ...  }&\multicolumn{1}{c}{ ...  }&\multicolumn{1}{c}{ ... } \\
            &(0.28)&(0.05)&(0.17)&(0.12)&(0.70)&&(...) &(1.06)&&\multicolumn{1}{c}{(...) }&\multicolumn{1}{c}{(...) }&\multicolumn{1}{c}{(...) }&\multicolumn{1}{c}{(...)} \\
J0837$+$1921& 22.10& 21.91& 22.87& 21.99& 21.84&& 21.25& 21.53&&\multicolumn{1}{c}{ ...  }&\multicolumn{1}{c}{ ...  }&\multicolumn{1}{c}{ ...  }&\multicolumn{1}{c}{ ... } \\
            &(0.14)&(0.05)&(0.17)&(0.13)&(0.38)&&(0.33)&(0.49)&&\multicolumn{1}{c}{(...) }&\multicolumn{1}{c}{(...)} &\multicolumn{1}{c}{(...) }&\multicolumn{1}{c}{(...)} \\
J1004$+$3256& 22.13& 21.33& 22.13& 20.98& 22.24&& ...  & 22.35&&\multicolumn{1}{c}{ ...  }&\multicolumn{1}{c}{ ...  }&\multicolumn{1}{c}{ ...  }&\multicolumn{1}{c}{ ... } \\
            &(0.23)&(0.05)&(0.16)&(0.09)&(0.81)&&(...) &(0.18)&&\multicolumn{1}{c}{(...) }&\multicolumn{1}{c}{(...)} &\multicolumn{1}{c}{(...) }&\multicolumn{1}{c}{(...)} \\
J1206$+$5007& 22.21& 21.76& 22.42& 22.24& 21.42&& 22.28& 22.25&&\multicolumn{1}{c}{ 17.13}&\multicolumn{1}{c}{ 17.10}&\multicolumn{1}{c}{ ...  }&\multicolumn{1}{c}{ ... } \\
            &(0.29)&(0.07)&(0.20)&(0.25)&(0.47)&&(0.14)&(0.12)&&\multicolumn{1}{c}{(0.09)}&\multicolumn{1}{c}{(0.30)}&\multicolumn{1}{c}{(...) }&\multicolumn{1}{c}{(...)} \\
J1234$+$3901& 21.97& 21.92& 21.93& 21.47& 22.24&& 21.37& 21.40&&\multicolumn{1}{c}{ ...  }&\multicolumn{1}{c}{ ...  }&\multicolumn{1}{c}{ ...  }&\multicolumn{1}{c}{ ... } \\
            &(0.14)&(0.06)&(0.09)&(0.09)&(0.54)&&(0.08)&(0.07)&&\multicolumn{1}{c}{(...) }&\multicolumn{1}{c}{(...) }&\multicolumn{1}{c}{(...) }&\multicolumn{1}{c}{(...)} \\
J1505$+$3721& 21.20& 20.97& 21.69& 20.60& 21.79&& 21.54& 21.29&&\multicolumn{1}{c}{ ...  }&\multicolumn{1}{c}{ ...  }&\multicolumn{1}{c}{ ...  }&\multicolumn{1}{c}{ ... } \\
            &(0.11)&(0.03)&(0.08)&(0.04)&(0.48)&&(0.32)&(0.18)&&\multicolumn{1}{c}{(...) }&\multicolumn{1}{c}{(...) }&\multicolumn{1}{c}{(...) }&\multicolumn{1}{c}{(...)} \\
J2229$+$2725& 21.96& 21.47& 22.27& 20.97& 21.43&& 21.45& 21.88&&\multicolumn{1}{c}{ ...  }&\multicolumn{1}{c}{ ...  }&\multicolumn{1}{c}{ ...  }&\multicolumn{1}{c}{ ... } \\
            &(0.15)&(0.04)&(0.11)&(0.05)&(0.33)&&(0.28)&(0.37)&&\multicolumn{1}{c}{(...) }&\multicolumn{1}{c}{(...) }&\multicolumn{1}{c}{(...) }&\multicolumn{1}{c}{(...)} \\
\hline
\end{tabular}



  \end{table*}

Very low-metallicity galaxies are much fainter than the LyC leaking
galaxies studied before at $z$ $\sim$ 0.3 -- 0.4. In fact, no galaxy with 12+log(O/H)
$\la$ 7.25 is known in the Sloan Digital Sky Survey (SDSS) at these redshifts. 
Only at very low redshifts, $z$ $\la$ 0.1, are the extremely low-metallicity galaxies sufficiently
bright to be selected for spectroscopic observations in the SDSS.
 Since these XMDs have low redshifts, direct observations of
the Lyman continuum are not possible with the {\sl HST}/COS. 
Therefore, our intention is to use the Ly$\alpha$
profiles in their spectra to derive information 
about possible leaking LyC radiation in galaxies that have much
lower stellar masses and metallicities than confirmed LyC leakers. 

The primary objective of this paper is to obtain for the first time UV spectra
with the Ly$\alpha$ line of nine low-redshift ($z$ $\la$ 0.13) compact,
extremely metal-deficient (XMD) SFGs, with 12+logO/H = 6.97 - 7.23.
Such galaxies are extremely rare
in the local Universe, but they likely were common at the epoch of
reionization \citep*[e.g. ][]{La20}.
The proposed XMD sample is the most complete one known so far, including all
objects gathered from the latest Data Release 16 (DR16) of the SDSS, that
satisfy the selection criteria
described below. 
This sample
includes the two SFGs with the lowest 12+log(O/H) known until now,
J0811$+$4730 and J1234$+$3901 \citep[12+log(O/H)=6.98, 7.03, ][]{I18c,I19}.
To date, only one {\sl HST}/COS spectrum has been obtained for a SFG in that
extreme metallicity range, but with much less extreme conditions in its
H~{\sc ii} regions, that of I~Zw~18 \citep{Ja14}. The UV spectrum of~I Zw~18 
shows a broad Ly$\alpha$ absorption line, 
without any sign of Ly$\alpha$ emission.

The proposed observations will considerably
increase the sample of XMDs observed in the Ly$\alpha$ line range, and extend
the range of 12+log(O/H) down to O/H $\sim$1/50 of the solar value, as derived by
\citet{A09}. Because of their very low metallicities, low stellar masses and
extremely high star-formation activity, these XMDs are likely the closest
local analogues of the primordial dwarf galaxies in the early Universe.
Strong Ly$\alpha$ in
emission has been seen in most low-$z$ LyC leakers with higher metallicity by
\citet{I16,I16b,I18,I18b,I21a,I22} and \citet{Fl22}.
The galaxies studied here have lower metallicities and stellar masses than previously
observed Ly$\alpha$ emitting galaxies such as the galaxies at $z$ $<$ 0.07 with
extreme O$_{32}$ ratios of $>$~20 \citep{I20}, the ``green pea'' (GP) galaxies
with $z$~$\sim$~0.1~--~0.3
studied e.g. by \citet{J17}, \citet{Y17}, \citet{MK19}, and the confirmed LyC 
leakers at $z$~$\sim$~0.3~--~0.4 \citep{I16,I16b,I18,I18b}.

We examine here whether the lowest-metallicity galaxies also possess strong
Ly$\alpha$ emission. Such objects will be called hereafter ``extreme Ly$\alpha$ emitters'' or ``xLAEs''. The observations probe a new
metallicity domain with {\sl HST}, and the spectra provide important
insight on a variety of topics/questions concerning these galaxies, including their Ly$\alpha$ and LyC escape,
their ISM and radiation field.

  \begin{table*}
  \caption{Integrated characteristics \label{tab3}}
\begin{tabular}{lccccccrccccc} \hline
Name&$M_{\rm FUV}^{\rm SED}$$^{\rm a}$&$M_{\rm FUV}$$^{\rm b}$&  $M_g$$^{\rm c}$   &log $M_\star$$^{\rm d}$ &log $L$(H$\beta$)$^{\rm e}$&SFR$^{\rm f}$&sSFR$^{\rm g}$&log $\xi_{\rm ion}$$^{\rm h}$&$\alpha$$^{\rm i}$&$r_{50}$$^{\rm j}$&$\Sigma_1$$^{\rm k}$&$\Sigma_2$$^{\rm l}$\\
\hline   
J0122$+$0048&$-$16.15&$-$15.16&$-$15.44&7.00&40.02&0.23, 0.05& 23&25.45&0.12&0.030&~\,5.1&~\,81\\
J0139$+$1542&$-$14.43&$-$11.59&$-$13.63&5.15&39.47&0.06, 0.02&425&25.54&0.04&0.013&11.9  &113\\
J0811$+$4730&$-$15.21&   ...  &$-$15.10&5.88&39.94&0.19, 0.06&250&25.71& ...& ... & ...  &...\\
J0837$+$1921&$-$16.12&$-$16.17&$-$15.51&6.24&40.10&0.28, 0.08&161&25.53&0.14&0.023&~\,4.5&169\\
J1004$+$3256&$-$16.42&   ...  &$-$16.09&6.24&40.31&0.45, 0.13&259&25.67&0.14&0.030&~\,7.3&159\\
J1206$+$5007&$-$15.52&$-$14.53&$-$15.05&6.26&39.82&0.15, 0.05& 82&25.46&0.13&0.030&~\,2.8&~\,53\\
J1234$+$3901&$-$18.03&$-$17.61&$-$17.08&6.94&40.85&1.57, 0.56&180&25.74&0.21&0.050&11.3  &200\\
J1505$+$3721&$-$16.71&$-$16.14&$-$16.71&6.48&40.70&1.11, 0.29&368&25.69&0.19&0.045&~\,9.8&175\\
J2229$+$2725&$-$16.43&$-$16.25&$-$16.23&6.39&40.49&0.68, 0.08&640&25.81&0.20&0.040&12.5  &313\\
\hline
  \end{tabular}

\hbox{$^{\rm a}$$M_{\rm FUV}^{\rm SED}$ is the absolute FUV magnitude derived from the intrinsic rest-frame SED.}

\hbox{$^{\rm b}$$M_{\rm FUV}$ is the absolute FUV magnitude derived from the apparent {\sl GALEX} magnitude.}

\hbox{$^{\rm c}$$M_g$$^{\rm c}$ is the absolute SDSS $g$ magnitude corrected for the Milky Way extinction.}

\hbox{$^{\rm d}$$M_\star$ is the stellar mass in solar masses.}

\hbox{$^{\rm e}$$L$(H$\beta$) is the extinction-corrected H$\beta$ luminosity 
in erg s$^{-1}$.}

\hbox{$^{\rm f}$SFR is the star-formation rate in M$_\odot$ yr$^{-1}$. The first
value is derived from the extinction-corrected H$\beta$ luminosity whereas the second}

\hbox{~value is
determined from the extinction-corrected UV luminosity at $\lambda$ = 1500\AA.}

\hbox{$^{\rm g}$Specific star formation rate sSFR=SFR/$M_\star$ in Gyr$^{-1}$,
where SFR is derived from the H$\beta$ luminosity.}

\hbox{$^{\rm h}$$\xi_{\rm ion}$ is the ionizing photon production efficiency 
in Hz erg$^{-1}$ defined as $\xi_{\rm ion}$ = $N_{\rm LyC}$/$L_\nu$, where 
$N_{\rm LyC}$ is the Lyman continuum}

\hbox{~photon production rate derived from the extinction-corrected
H$\beta$ luminosity and $L_\nu$ is the monochromatic luminosity}

\hbox{~at $\lambda$ = 1500\AA\ derived from the intrinsic rest-frame SED.}

\hbox{$^{\rm i}$$\alpha$ is the exponential disc scale length in kpc.}

\hbox{$^{\rm j}$$r_{50}$ is the galaxy radius in kpc where the NUV intensity is equal to half of the maximal intensity.}

\hbox{$^{\rm k}$$\Sigma_1$ is the star-formation rate surface density in M$_\odot$yr$^{-1}$kpc$^{-2}$, assuming the galaxy radius to be equal to $\alpha$.}

\hbox{$^{\rm l}$$\Sigma_2$ is the star-formation rate surface density in M$_\odot$yr$^{-1}$kpc$^{-2}$, assuming the galaxy radius to be equal to $r_{50}$.}

  \end{table*}

\begin{table}
\caption{References on data for $z$ $\ga$ 6 galaxies  \label{tab4}}
\begin{tabular}{lll} \hline
Figure&Relation&References\\ \hline
Fig.~\ref{fig1}b&O$_{32}$--R$_{23}$             &\citet{Sa23}\\
                &&\citet{He22}\\
Fig.~\ref{fig1}c&12+log(O/H)--$M_\star$/M$_\odot$&\citet{La22}\\
                &&\citet{Jo20}\\
                &&\citet{He22}\\
                &&\citet{A23}\\
Fig.~\ref{fig1}d&O$_{32}$--$M_\star$/M$_\odot$&\citet{He22}\\
Fig.~\ref{fig1}e&log $\xi_{\rm ion}$--12+log(O/H)&\citet{A23}\\
Fig.~\ref{fig1}f&log $\xi_{\rm ion}$--O$_{32}$&\citet{Sa23}\\
Fig.~\ref{fig6}a&$L$(Ly$\alpha$)--M$_{\rm FUV}$&\citet{Ju22}\\
                &&\citet{Ni23}\\
                &&\citet{Fu20}\\
Fig.~\ref{fig6}b&EW(Ly$\alpha$)--M$_{\rm FUV}$&\citet{Ju22}\\
                &&\citet{Ni23}\\
                &&\citet{Jo20}\\
                &&\citet{Fu20}\\
                &&\citet{Sa23}\\
                &&\citet{Pe18}\\
Fig.~\ref{fig7}a&$f_{\rm esc}$(Ly$\alpha$)--M$_{\rm FUV}$&\citet{Sa23}\\
                &&\citet{Ni23}\\
Fig.~\ref{fig7}c&$f_{\rm esc}$(Ly$\alpha$)--O$_{32}$&\citet{Sa23}\\
Fig.~\ref{fig7}e&$f_{\rm esc}$(Ly$\alpha$)--$L$(Ly$\alpha$)&\citet{Si23}\\
                &&\citet{Ni23}\\
Fig.~\ref{fig7}f&$f_{\rm esc}$(Ly$\alpha$)--log $\xi_{\rm ion}$&\citet{Sa23}\\
                &&\citet{Ni23}\\
                &&\citet{Si23}\\
Fig.~\ref{fig7}g&$f_{\rm esc}$(Ly$\alpha$)--EW(Ly$\alpha$)&\citet{Sa23}\\
                &&\citet{Ni23}\\
\hline
\end{tabular}
\end{table}

\begin{figure*}
\includegraphics[angle=0,width=0.90\linewidth]{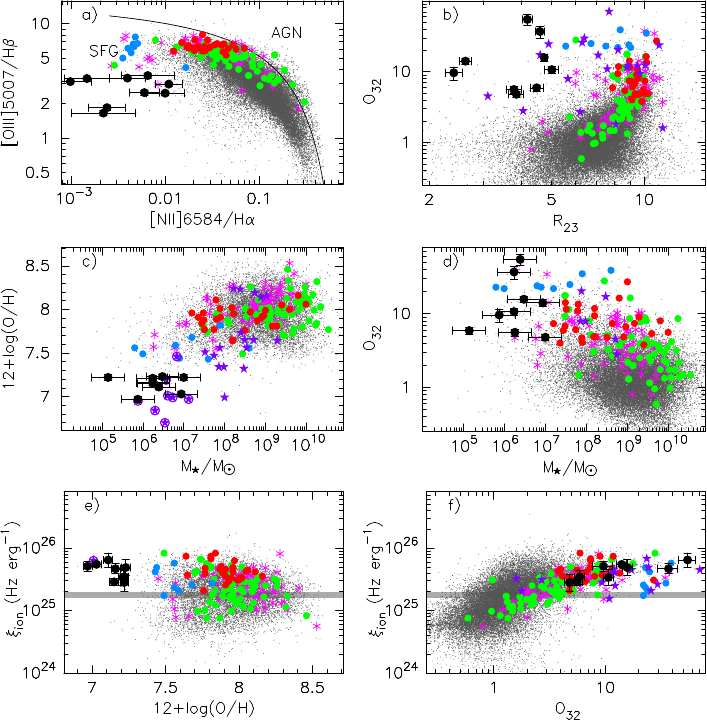}
\caption{{\bf a)} The Baldwin-Phillips-Terlevich (BPT) diagram \citep{BPT81} for
SFGs. {\bf b)} The O$_{32}$ -- R$_{23}$ diagram for SFGs where R$_{23}$ = 
([O~{\sc ii}]3727 + [O~{\sc iii}]4959 + [O~{\sc iii}]5007)/H$\beta$.
{\bf c)} and {\bf d)} Dependencies of the oxygen abundances 12+log(O/H) and
O$_{32}$ ratios, respectively, on the stellar masses $M_\star$.
{\bf e)} and {\bf f)} Dependencies of the ionizing photon efficiencies
$\xi_{\rm ion}$ on the oxygen abundances 12+log(O/H) and O$_{32}$ ratios,
respectively.
In all panels, the xLAEs (this paper) with the respective errors
are shown by black filled circles,
the low-redshift ($z$ $<$ 0.1) LAEs with high O$_{32}$
\citep{I20} are represented by blue filled circles, LyC leaking galaxies by 
\citet{I16,I16b,I18,I18b,I21a,I22} are shown by red filled circles, 
the LyC galaxies by \citet{Fl22} are shown by green filled circles and
low-$z$ LAEs by \citet{H15}, \citet{JO14}, \citet{J17}, \citet{Y16,Y17},
\citet{Xu22}, \citet{Hu23} are represented by magenta asterisks. The references
on the high-$z$ ($z$ $>$ 6) LAEs (purple stars) are given in Table \ref{tab4}.
Separately, by encircled purple stars, are
shown $z$ $\sim$ 6 -- 7 XMDs in {\bf c)} and their $\xi_{\rm ion}$ and
12+log(O/H) average values in {\bf e)} \citep{A23}.
The compact SFGs from the SDSS \citep{I21b} are represented by grey dots.
The solid line in {\bf a)} by \citet{K03} separates SFGs from active galactic 
nuclei (AGN). Values of $\xi_{\rm ion}$ assumed in canonical Universe
reionisation models are shown in {\bf e)} and {\bf f)} with a horizontal grey
shaded region \citep[e.g. ][]{R13,Bo16}.
\label{fig1}}
\end{figure*}

The selection criteria are presented in Section~\ref{sec:select}.
The properties of the selected galaxies derived from SDSS
observations in the optical range, and from UV and mid-infrared observations
with the {\sl GALEX} and {\sl WISE} space telescopes, respectively, are
discussed in Section~\ref{sec:integr}.
The {\sl HST} observations and data reduction are described
in Section~\ref{sec:obs}. The surface brightness profiles in the UV range are 
discussed in Section~\ref{sec:sbp}. Ly$\alpha$ emission is considered in 
Section~\ref{sec:lya}.  In Section~\ref{discussion}, we compare
the Ly$\alpha$ escape fractions obtained for our galaxies in this paper and
for other galaxies in some recent studies. Our results are summarized in Section~\ref{summary}.

\section{Selection criteria}\label{sec:select}

We use the SDSS Data Release 16 (DR16) \citep{A20} to select galaxies, applying
the following criteria:

- the temperature-sensitive [O~{\sc iii}] $\lambda$4363 emission line 
in the SDSS spectrum is detected at the level greater than 3$\sigma$ to
allow accurate abundance determination;

- the oxygen abundance of the ionized gas in the H\,{\sc ii} region, derived
using the direct $T_{\rm e}$-method, is 12+log(O/H)~$<$~7.25;

- the redshift $z$ of the galaxy is greater than 0.02, to avoid contamination
from the wings of the Milky Way Ly$\alpha$ absorption line;

- no bright objects must be present in vicinity of the galaxy to
satisfy COS safety requirements;

- the galaxy is sufficiently compact in the SDSS image for detection of the
total galaxy flux inside the 2.5 arcsec (in diameter) COS spectroscopic
aperture;

- the galaxy should be isolated inside the 2.5 arcsec COS spectroscopic aperture
to avoid contamination from other sources;

- the {\sl GALEX} NUV magnitude of the galaxy is brighter than $\sim$ 22.5 mag
to be acquired with COS and to obtain spectra of reasonably good quality.

These selection criteria gave a total sample of nine galaxies. 

\section{Integrated properties of sample galaxies} \label{sec:integr}

The SDSS spectra of the selected galaxies reveal extremely low oxygen 
abundances, in the range 12~+~log(O/H)~=~6.97~--~7.23
(Tables~\ref{tab1}, \ref{taba2}). These are considerably lower as compared to the oxygen abundances 
of the low-redshift Ly$\alpha$ and LyC leakers observed so
far.

The sample galaxies show very high rest-frame equivalent widths EW(H$\beta$) =
156 -- 580~\AA\ (Table \ref{tab1}), indicating very recent star formation.
All galaxies are characterised by high O$_{32}$ ratios ranging from $\sim$5
to the very extreme value of $\sim$55. Both high values of EW(H$\beta$) and O$_{32}$ are
consequences of the selection criterion requiring the detection of the
[O~{\sc iii}] $\lambda$4363 emission line with sufficient accuracy
(Table~\ref{taba1}). This line is strongest in the
youngest bursts of star formation.

All selected galaxies are faint in the optical range, with the SDSS $g$
magnitude fainter than $\sim$21 mag (Table~\ref{tab2}). However, despite their
faintness in the optical range, all galaxies were detected by
{\sl Galaxy Evolution Explorer} ({\sl GALEX}) in the NUV range and
most of them were detected in the FUV range, because of a 
strongly rising continuum from the optical to the UV range. On the other
hand, only one galaxy was detected in the mid-infrared range by the
{\sl Wide-field Infrared Survey Explorer} ({\sl WISE}) (Table~\ref{tab2}),
indicating that emission from the red stellar population and dust is very weak.

The interstellar extinction is obtained from the hydrogen Balmer decrement in
the SDSS spectra. We derive the ionized gas metallicity from the
same spectra using the prescriptions of \citet*{ITL94} and \citet{I06}.

The emission-line fluxes corrected for extinction, adopting
the \citet*{C89} reddening law with $R(V)$ = 2.7, the extinction 
coefficients $C$(H$\beta$), the rest-frame equivalent 
widths of the  H$\beta$, [O~{\sc iii}]$\lambda$5007 and H$\alpha$ emission
lines, and the observed H$\beta$ fluxes are shown in Table \ref{taba1}.
The fluxes from Table \ref{taba1} and the direct $T_{\rm e}$ method are 
used to derive the electron temperatures, the electron number densities and the
element abundances in the H~{\sc ii} regions. These quantities are shown in
Table~\ref{taba2}. 

However, we note that oxygen abundances derived from the
SDSS spectra can sometime be incorrect because strong emission lines such as
[O~{\sc iii}] $\lambda$4959, 5007 can be clipped, resulting in overestimated
electron temperatures and, thus underestimated oxygen abundances. Therefore,
additional observations with other telescopes
are needed to check the very low metallicities of the selected galaxies.
We have used the LBT to obtain spectrophotometric observations
for the three galaxies with the lowest metallicities in our sample, J0811$+$4730
\citep{I18c}, J1234$+$3901 \citep*{I19}, and J2229$+$2725 \citep*{I21c}. The
SDSS abundances are in good agreement with the LBT ones for these three
galaxies.  

  \begin{table}
  \caption{{\sl HST}/COS observations \label{tab5}}
  \begin{tabular}{lccc} \hline
\multicolumn{1}{c}{Name}&\multicolumn{1}{c}{Date}&\multicolumn{2}{c}{Exposure time (s)} \\ 
&    &MIRRORA&G130M \\
&    &       &(Central\\
&    &       &wavelength) (\AA)\\
\hline
J0122$+$0048&2022-09-16&2$\times$400     & 6889\\
            &          &         &(1222)\\
J0139$+$1542&2022-11-14&2$\times$600      & 9337\\
            &          &         &(1222)\\
J0811$+$4730&2023-03-25&2$\times$0        & 5658\\
            &          &         &(1222)\\
J0837$+$1921&2023-03-20&2$\times$500      & 9537\\
            &          &         &(1222)\\ 
J1004$+$3256&2023-01-20&2$\times$600      & 9433\\
            &          &         &(1222)\\ 
J1206$+$5007&2023-03-09&2$\times$600      &10045\\
            &          &         &(1222)\\ 
J1234$+$3901&2022-11-16&2$\times$600      & 9526\\
            &          &         &(1291)\\ 
J1505$+$3721&2023-02-22&2$\times$400      & 7058\\
            &          &         &(1291)\\ 
J2229$+$2725&2022-09-20&2$\times$500      & 9586\\
            &          &         &(1291)\\ 
\hline
\end{tabular}
  \end{table}

The emission-line luminosities and stellar masses of our galaxies were obtained 
adopting a luminosity distance \citep[NASA Extragalactic Database
  (NED),][]{W06}, with the cosmological 
parameters $H_0$=67.1 km s$^{-1}$Mpc$^{-1}$, $\Omega_\Lambda$=0.682, 
$\Omega_m$=0.318 \citep{P14}. We also derived absolute magnitudes
$M_g$ and $M_{\rm FUV}$ in the SDSS $g$ and
{\sl GALEX} FUV bands (Table \ref{tab3}), assuming negligible dust extinction. 

The H$\beta$ luminosities $L$(H$\beta$) and corresponding star-formation rates 
SFR(H$\beta$) shown in Table \ref{tab3} were obtained from the extinction-corrected 
H$\beta$ fluxes, using the relation by \citet{K98} and adopting
the extinction-corrected H$\alpha$/H$\beta$ flux ratios. They are more than
one order of magnitude lower than the respective values for
LyC leakers \citep{I16,I16b,I18,I18b,I21a,I22}. Specific star formation rates
sSFR(H$\beta$) = SFR(H$\beta$)/$M_\star$ are $\sim$~100~Gyr$^{-1}$, similar to
sSFR(H$\beta$) for LyC leakers \citep{I16,I16b,I18,I18b,I21a}
and are several orders
of magnitude higher than the sSFRs of the SDSS main sequence SFGs.
We also use the extinction-corrected monochromatic UV luminosity at the
rest-frame wavelength of 1500\AA\ and equation 2 from \citet*{Ma98} to derive
SFR(UV). We find it to be several times lower than SFR(H$\beta$) (Table~\ref{tab3}).
A similar difference between SFR(H$\beta$) and SFR(UV) has also been found by other authors, 
for example 
by \citet{A23} for a sample of low-mass and low-metallicity SFGs at $z$~$\sim$~6~--~8.
This likely
is typical for low-mass galaxies with bursts of star formation as the relations
by \citet{K98} and \citet{Ma98} are obtained by assuming long periods of star
formation of $\sim$ 100 Myr and 1 Gyr, respectively, which may not be applicable 
for our xLAEs. However, we show these quantities for the sake of
comparison because they are commonly used in many studies of bursting SFGs.

We use the extinction-corrected SDSS spectra of xLAEs to derive 
the stellar and nebular spectral energy 
distributions (SED), stellar masses, starburst ages and the modelled intrinsic
absolute magnitudes $M_{\rm FUV}^{\rm SED}$ in the FUV range. The star-formation
history is approximated by a single instantaneous burst with an age $<$~10~Myr
and a continuous star formation with a constant star-formation rate and an age
$>$~10~Myr. All parameters such as stellar mass, starburst age and time interval
for continuous star formation are derived iteratively by minimizing
differences between the continua of the modelled SED and SDSS spectra.
The SED fitting method, using a two-component stellar+nebular model,
is described e.g. in \citet{I18}.
We find that the stellar masses and FUV luminosities of our galaxies are 
$\sim$ 2 orders of magnitude lower than those of the confirmed LyC leakers 
(Table \ref{tab3}). Their extinction-corrected absolute FUV magnitudes are 
in general considerably fainter than those of the faintest galaxies at
$z$~=~6~--~8 observed thus far with the {\sl JWST}
\citep*[e.g. ][]{Jo23,M23,Sa23,Sh23}.

In Fig.~\ref{fig1} we compare the integrated properties of our selected
galaxies with various samples of low-$z$ LyC leaking galaxies and LAEs
published by \citet{I16,I16b,I18,I18b,I20,I21a,I22},
\citet{Fl22}, \citet{JO14}, \citet{H15}, \citet{J17}, \citet{Y16,Y17},
\citet{Hu23}, as well as with the high-$z$ LAEs observed at
$z$ $\ga$ 6, during the period of the reionization. References for
high-$z$ galaxies are given in Table~\ref{tab4}. The errors are not
available for some galaxies in the comparison sample. Therefore, the error bars
in this Figure and all others are shown only for the galaxies in our sample
(black symbols).

In the Baldwin-Phillips-Terlevich
(BPT) diagnostic diagram (Fig. \ref{fig1}a), our sample galaxies (black filled circles) strongly deviate from the SFG branch
(grey dots), from low-redshift Ly$\alpha$ emitters with O$_{32}$ $\ga$ 20, but with
higher metallicities (blue filled circles), from $z$ $\sim$ 0.3 -- 0.4 LyC
leakers (red and green filled circles) and from other LAEs (magenta asterisks).
These deviations result from their extremely low metallicities. 
In this sense, our xLAEs are considerably more extreme than
other LAEs observed thus far with the {\sl HST}/COS.

In the same vein, it can be seen in Fig.~\ref{fig1}b, which shows the 
O$_{32}$ -- R$_{23}$ diagram, that the xLAEs
are also the most deviating objects from the SDSS main SFG sequence (grey dots),
again because of their extremely low metallicities. On the other
hand, most of the confirmed LyC leakers with higher metallicities (red and green
filled circles) and other low-$z$ LAEs (magenta
asterisks) are mainly located
on the main sequence defined by SFGs. However, we
note that high-$z$ ($z$~$\ga$~6) LAEs
(purple stars in the Figure) are also systematically offset from the
main sequence, with the most deviating being the $z$ $\sim$ 8 -- 9 galaxies of
\citet{He22}, indicating their extremely low metallicities.

Our xLAEs follow the mass-metallicity relation (Fig.~\ref{fig1}c),
having until recently  both lower metallicities and masses than
low-$z$ LyC leakers and 
high-$z$ LAEs (compare the locations of the black circles with
those of the red and green circles and purple stars in Fig.~\ref{fig1}c).
But  \citet{A23} have recently discussed a sample of $z$ $\sim$ 6 -- 8 XMDs
with stellar masses and oxygen abundances similar to those of the xLAEs.
These galaxies are represented by encircled purple stars in Fig.~\ref{fig1}c).
We note, however, that only the H$\beta$ and [O~{\sc iii}] $\lambda$4959, 5007
emission lines were detected in most of these galaxies, whereas only upper
limits were obtained for the [O~{\sc ii}] $\lambda$3727 emission line
intensities. Therefore, the oxygen abundance is derived from
the [O~{\sc iii}] $\lambda$5007/H$\beta$ flux ratio, which depends not only
on the oxygen abundance, but also on the ionization parameter. Therefore, the
oxygen abundances derived by \citet{A23} are somewhat uncertain. We also note
that no data on Ly$\alpha$ emission are presented.

Our xLAEs also have on
average higher O$_{32}$ ratios compared e.g. to those of low-$z$ LyC leakers and
high-$z$ LAEs (compare the locations of the black circles with
those of the red, green and purple symbols in Fig.~\ref{fig1}d).

Finally, all xLAEs have very high ionizing photon production
efficiencies $\xi_{\rm ion}$ (Fig.~\ref{fig1}e). Their values compare well with
those for the high-$z$ XMDs of \citet{A23},
low-$z$ LAEs with high O$_{32}$ ratios of $>$ 20
\citep{I20}, and for low-$z$ LyC leaking galaxies from the
\citet{I16,I16b,I18,I18b,I21a,I22} sample.
This is because all have high EW(H$\beta$) and thus young starburst ages.
The $\xi_{\rm ion}$ values of our xLAEs are well above the canonical
values (horizontal lines) usually used in the Universe
reionization models \citep{Bo16}. For comparison, LyC leaking galaxies from
the \citet{Fl22} sample (green filled circles) and low-$z$ LAEs
(magenta asterisks) have lower $\xi_{\rm ion}$s, which compare well with the
canonical value and those of the SDSS main sequence star-forming galaxies. We also
note that there is no dependence of $\xi_{\rm ion}$ on the oxygen abundance
12+log(O/H) (Fig.~\ref{fig1}e), whereas there is a clear dependence of $\xi_{\rm ion}$ on the
O$_{32}$ ratio (Fig.~\ref{fig1}f). This is a consequence of the direct
dependence of O$_{32}$ on the ionization parameter.

\begin{figure*}
\includegraphics[angle=0,width=0.90\linewidth]{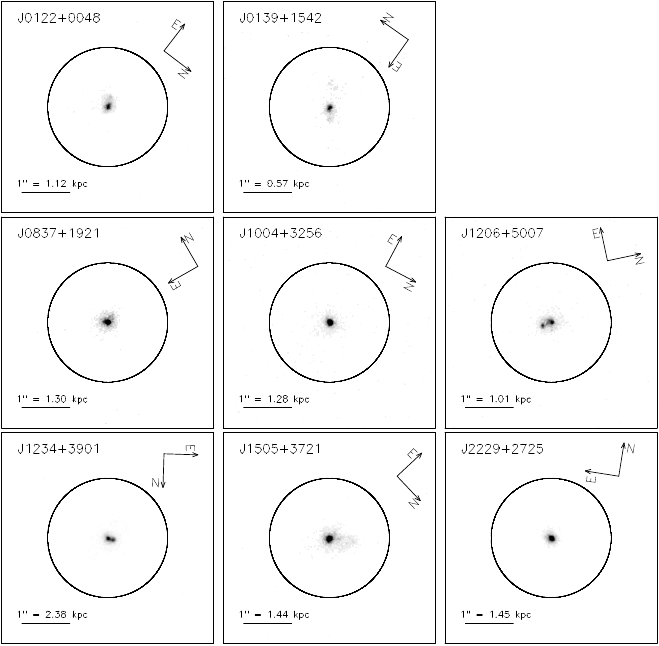}
\caption{The {\sl HST}/COS NUV acquisition images of the most metal-deficient
galaxies in log surface brightness scale. The image for J0811$+$4730 is not
shown (see text). A circle in all panels is the COS spectroscopic aperture with
a diameter of 2.5 arcsec. The linear scale in each panel is derived adopting the
angular size distance (Table~\ref{tab1}).
\label{fig2}}
\end{figure*}

\begin{figure*}
\includegraphics[angle=0,width=0.90\linewidth]{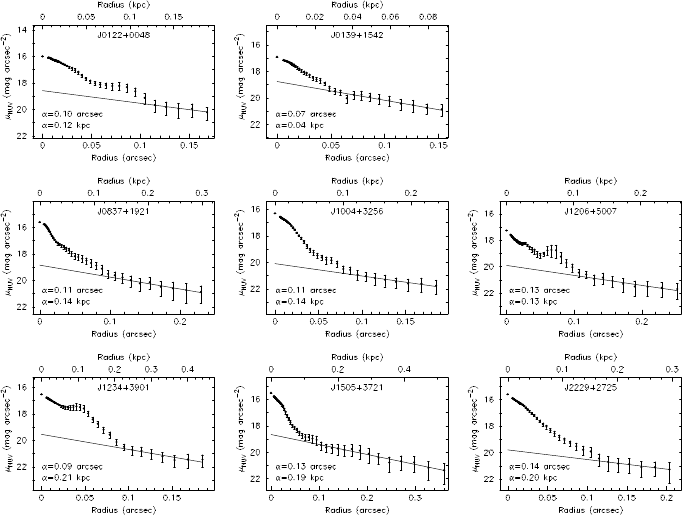}
\caption{NUV surface brightness profiles of the most metal-deficient galaxies.
The profile for J0811$+$4730 is not shown. Solid lines represent the linear fits
to the surface brightness profiles of the outer galaxy parts.
\label{fig3}}
\end{figure*}

\section{{\sl HST}/COS observations and data 
reduction}\label{sec:obs}

{\sl HST}/COS observations of the nine most metal-deficient galaxies were
obtained during the period September 2022 -- March 2023 in program GO~16672
(PI: Y.\ I.\ Izotov), (Table \ref{tab5}). The galaxies were acquired by COS
near-ultraviolet (NUV) imaging. The brightest region in the image of each target
was centered in the
2.5\,arcsec diameter spectroscopic aperture (Fig. \ref{fig2}). Unfortunately,
acquisition observations for the most metal-deficient galaxy in our
sample, also the most metal-deficient SFG known to-date, J0811$+$4730, 
failed. Therefore, the imaging data for this galaxy are not present.
The spectroscopic observations of J0811$+$4730 were also affected
by technical problems, so only part of the scheduled time was used.
However, the quality of the J0811$+$4730 spectrum was sufficiently high so that
we decided not to request a repeat observation.

The other eight galaxies show a very compact structure (Fig.~\ref{fig2}),
superposed on a 
low-surface-brightness (LSB) component. Most of the NUV continuum in these
galaxies is concentrated in central regions with full widths at half maximum
$\la$ 0.1 arcsec, which is  much smaller than the central unvignetted
$0.8$\,arcsec diameter region of the spectroscopic aperture.

Spectra with Ly$\alpha$ emission lines were obtained
with the COS G130M grating, positioned at the central wavelength of 1222\,\AA\
for six galaxies, and at 1291\,\AA\ for three galaxies
with higher redshifts (J1234$+$3901, J1505$+$3721 and J2229$+$2725) because
the resolving power of the 1222\AA\ setup declines at
wavelengths longer than ~$\sim$ 1250\AA.
The grating was positioned at the COS Detector 
Lifetime Position 4, yielding a spectral resolving power 
$\lambda/\Delta\lambda\simeq 14,000$ in the wavelength range of interest.
Spectra with the grating centered at 1222\,\AA\ were obtained in Segments A
and B, covering a wavelength range of $\sim$~300\,\AA, whereas spectra with
the grating centered at 1291\,\AA\ were obtained only in Segment A, giving a
wavelength coverage of $\sim$~150\,\AA. No observation with Segment B is allowed for the latter configuration. All 
four focal-plane offset positions were employed to correct for grid wire shadows
and detector blemishes. For two galaxies, J1505$+$3721 and J2229+2725, the
blue components of Ly$\alpha$ emission are partly contaminated by emission
from the geocoronal O~{\sc i} $\lambda$1304 line. In those cases, a special
requirement was made to observe those galaxies only during the periods with
maximal shadow time intervals, when the brightness of the geocoronal line is
minimal. Fortunately, the effect of contamination was not high, allowing to
obtain the profiles of the Ly$\alpha$ blue components in both galaxies.

The spectra were reduced with the CALCOS pipeline v3.3.4, followed 
by accurate background subtraction and co-addition of individual exposures with
custom software \citep{Ma21}. These reduction procedures are analogous to those
used by \citet{I16,I16b,I18,I18b,I20,I21a,I22}.

\begin{figure*}
\includegraphics[angle=0,width=0.90\linewidth]{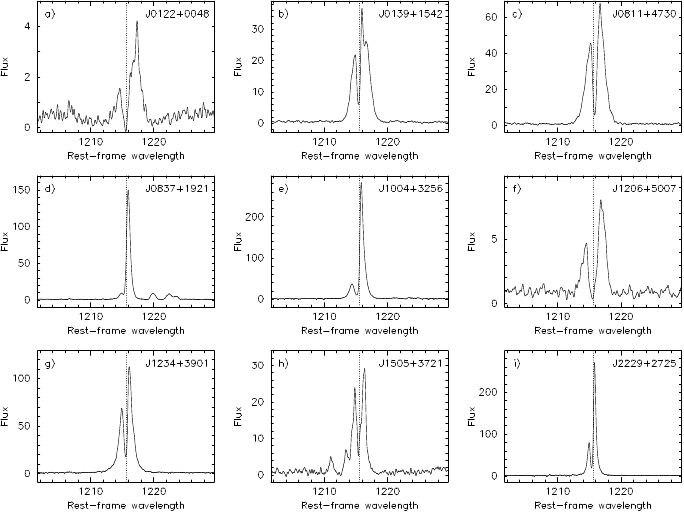}
\caption{Ly$\alpha$ profiles. Vertical dotted lines indicate the centres of 
profiles. The geocoronal O~{\sc i} $\lambda$1301,1304\AA\ emission in spectra
was removed by considering only orbital night data in the affected wavelength
range. The removal of these lines is most important for J1505$+$3721 and
J2229$+$2725 as they are very close to or overlapping with the Ly$\alpha$
emission line. Flux densities are in 10$^{-16}$ erg s$^{-1}$ cm$^{-2}$\AA$^{-1}$ 
and wavelengths are in \AA. \label{fig4}}
\end{figure*}

\begin{figure*}
\includegraphics[angle=0,width=0.90\linewidth]{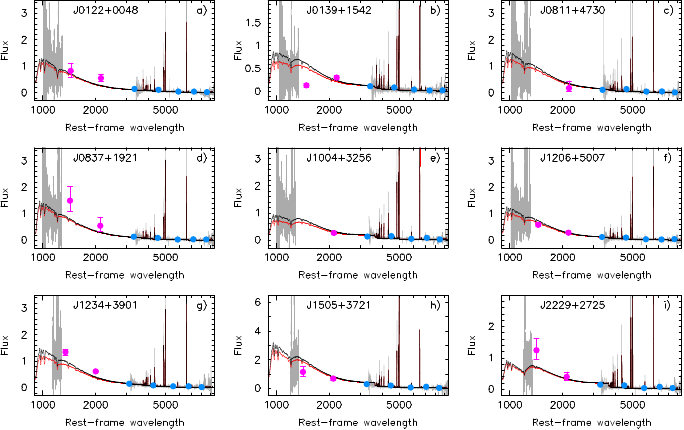}
\caption{A comparison of the COS G130M and SDSS spectra (grey lines), and
photometric data with the modelled SEDs. Photometric data
for {\sl GALEX} FUV and NUV ﬂuxes with 1$\sigma$ error bars (magenta symbols)
and SDSS ﬂuxes in $u, g, r, i, z$ bands (blue symbols)
are shown. Modelled intrinsic SEDs, which are reddened by the
Milky Way extinction using \citet{C89} reddening law with $R(V)_{\rm MW}$ = 3.1
and internal extinction with
$R(V)_{\rm int}$ = 3.1 and 2.7, are shown by black and red solid lines,
respectively. Flux densities are in 10$^{-16}$ erg s$^{-1}$ cm$^{-2}$ \AA$^{-1}$,
wavelengths are in \AA.
\label{fig5}}
\end{figure*}

\section{Surface brightness distribution in the NUV range}\label{sec:sbp}

Using the COS acquisition images, 
we obtain the surface brightness (SB) profiles of xLAEs in the NUV range
with a linear decrease (in magnitudes) in the outer part and a sharp increase
in the central part because of the presence of the bright star-forming region(s)
(Fig.~\ref{fig3}). The shape of the SB profiles in these galaxies is similar to
those of confirmed LyC leakers \citep{I16,I16b,I18,I18b,I21a,I22}, but with
the considerably smaller scale lengths $\alpha$ of our galaxies, in the range 
$\leq$ 0.21 kpc (Fig.~\ref{fig3}, Table~\ref{tab3}), compared to
$\alpha$ = 0.6 -- 1.8 kpc in LyC leakers \citep{I16,I16b,I18,I18b,I21a,I22}.
These scale lengths are similar to the smallest $\alpha$ found for
low- and high-$z$ compact star-forming galaxies \citep{P02,I20,Ma23}.
On the other hand, the surface densities of star-formation rate
$\Sigma$ = SFR(H$\beta$)/($\pi \alpha^2$) in the studied galaxies are
comparable to those of LyC leakers. 
The half-light radii $r_{50}$ of our galaxies in the NUV are 
considerably smaller than $\alpha$ because of the compactness of the bright
star-forming regions (see Table~\ref{tab3}).
Therefore, the corresponding surface densities of star-formation rate
$\Sigma$ = SFR(H$\beta$)/($\pi r_{50}^2$) are typically more than one order of
magnitude larger, and comparable to those found for 
low-redshift LyC leakers and SFGs at high redshifts \citep{CL16,PA18,Bo17}.

\section{Ly$\alpha$ emission}\label{sec:lya}

The rest-frame Ly$\alpha$ profiles in the medium-resolution spectra of the 9
xLAEs are shown in Fig. \ref{fig4}.
A double-peaked Ly$\alpha$ emission line is detected in the spectra
of all galaxies. In two galaxies, J0122$+$0048 and J1206$+$5007, the line is
weak and in J0122$+$0048, it is superposed on a broad absorption line.
In the remaining seven xLAEs, a strong Ly$\alpha$ $\lambda$1216\,\AA\
emission-line is seen, with the most intense one in J2229$+$2725.
The two-peak shape is similar to that observed in known LyC 
leakers \citep{I16,I16b,I18,I18b,V17} and in GP galaxies 
\citep{JO14,H15,Y17,MK19}. Some parameters of the Ly$\alpha$ emission line
such as flux, luminosity, equivalent width and separation between the
peaks are presented in Table~\ref{tab6}.

One of the goals of this paper is to derive for each galaxy the escape fraction of the
Ly$\alpha$ emission, and indirectly the absolute escape fraction of the Lyman
continuum. To achieve such a goal, some characteristics derived from the optical spectra
should be used, e.g. the fluxes of the H$\beta$ emission line. However,
observations in the UV and optical ranges were obtained with different
telescopes and somewhat different spectroscopic apertures. Therefore,
for the determination of the Ly$\alpha$ and LyC escape fractions, the
consistency of the fluxes in the UV and optical ranges should be checked. 
One approach would be to see how well the extrapolation toward UV wavelengths of the
optical spectral energy distribution from the SDSS spectrum would fit 
the UV COS spectrum.
Such a method has been 
adopted e.g. by \citet{I18,I18b,I20,I21a,I22} and we will use it here.

In Fig.~\ref{fig5}, we show the observed COS and SDSS spectra of our sample 
galaxies (grey lines), together with the {\sl GALEX} FUV and NUV (magenta filled circles with 1$\sigma$ errors)
and SDSS $u,g,r,i,z$ apparent magnitudes (blue filled circles). These photometric measurements are superposed on the reddened modelled SEDs of the optical
spectra and the SED extrapolations to the UV range. For the reddened SEDs
we adopt the extinction coefficients derived from the observed hydrogen Balmer
decrements and the reddening laws by \citet{C89} with $R(V)$ = 2.7 and 3.1 (red and black
lines, respectively, in Fig.~\ref{fig5}).

Several remarks can be made. First, we note that the SDSS spectra
are consistent with the total SDSS magnitudes, indicating that our galaxies
are very compact and that all the galaxy light falls into the
spectroscopic SDSS aperture. Therefore, aperture corrections are not needed.
Second, the red and black lines are nearly coincident because of the small
extinction. Third, the reddened SEDs satisfactorily reproduce both the observed
UV and optical spectra. Therefore, we can directly use extinction-corrected
Ly$\alpha$ and H$\beta$ fluxes to derive the Ly$\alpha$ escape fraction,
without any additional correction to adjust the UV COS spectrum to the extrapolation
of the optical SED. On the other hand, we note that the flux densities derived
from the {\sl GALEX} FUV and NUV magnitudes (magenta symbols in Fig.~\ref{fig5})
are poorly consistent with the flux densities of the COS spectra. Similar
discrepancies were noted earlier by \citet{I16b,I18,I18b,I21a,I22} in low-$z$
LyC galaxies. This indicates that {\sl GALEX} FUV and NUV magnitudes 
should not be used in general for accurate SED fitting, at least for faint galaxies with
COS observations.

  \begin{table*}
  \caption{Parameters for the Ly$\alpha$ and H$\beta$ emission lines \label{tab6}}
  \begin{tabular}{rcrcrcrrc} \hline
Name&$A$(Ly$\alpha$)$_{\rm MW}$$^{\rm a}$&\multicolumn{1}{c}{$I$(Ly$\alpha$)$^{\rm b}$}&log $L$(Ly$\alpha$)$^{\rm c}$&\multicolumn{1}{c}{EW(Ly$\alpha$)$^{\rm d}$}&\multicolumn{1}{c}{$V_{\rm sep}$$^{\rm e}$}&\multicolumn{1}{c}{$I$(H$\beta$)$^{\rm f}$}
&$f_{\rm esc}$(Ly$\alpha$)$^{\rm g}$&$f_{\rm esc}$(LyC)$^{\rm h}$\\ 
\hline
J0122$+$0048&0.291&  10.6$\pm$4.9&39.93$\pm$0.17&   14.1$\pm$5.5& 671.2$\pm$51.3  & 8.7$\pm$0.5& 5.2$\pm$~\,2.4& 1.0$\pm$0.5\\ 
J0139$+$1542&0.500& 121.4$\pm$5.3&40.35$\pm$0.02&  171.2$\pm$7.5& 252.0$\pm$26.1  & 7.8$\pm$0.5&66.8$\pm$~\,5.2&18.7$\pm$1.5~\,\\ 
J0811$+$4730&0.595& 272.4$\pm$7.7&41.11$\pm$0.01&  187.7$\pm$5.3& 365.5$\pm$22.6  &13.4$\pm$1.8&87.2$\pm$12.0  & 5.0$\pm$0.7\\ 
J0837$+$1921&0.220& 151.8$\pm$5.8&41.23$\pm$0.02&  131.6$\pm$5.0& 260.6$\pm$43.5  & 7.9$\pm$0.6&82.5$\pm$~\,7.0&16.8$\pm$1.4~\,\\ 
J1004$+$3256&0.138& 310.4$\pm$6.3&41.53$\pm$0.01&  239.9$\pm$4.9& 350.7$\pm$47.1  &25.3$\pm$1.1&52.7$\pm$~\,2.5& 5.8$\pm$0.3\\ 
J1206$+$5007&0.164&  17.1$\pm$5.9&40.03$\pm$0.13&   15.2$\pm$5.2& 548.3$\pm$59.0  & 9.9$\pm$0.7& 7.4$\pm$~\,2.6& 1.0$\pm$0.4\\ 
J1234$+$3901&0.123& 233.8$\pm$8.0&42.05$\pm$0.01&  174.8$\pm$6.0& 292.9$\pm$11.2  &17.3$\pm$1.0&58.0$\pm$~\,3.9&11.3$\pm$0.8~\,\\ 
J1505$+$3721&0.145&  53.4$\pm$9.9&40.88$\pm$0.07&   57.4$\pm$9.9& 369.9$\pm$17.8  &30.6$\pm$1.1& 7.5$\pm$~\,1.4& 4.7$\pm$0.9\\ 
J2229$+$2725&0.419& 301.8$\pm$7.8&41.64$\pm$0.01&  178.2$\pm$4.6& 207.0$\pm$16.2  &20.4$\pm$0.8&63.5$\pm$~\,3.0&34.2$\pm$1.6~\,\\ 
\hline
  \end{tabular}

\hbox{$^{\rm a}$$A$(Ly$\alpha$)$_{\rm MW}$ is the Milky Way extinction at the observed wavelength of the Ly$\alpha$
emission line in mags, adopting \citet{C89}}

\hbox{\, reddening law with $R(V)$=3.1.}

\hbox{$^{\rm b}$$I$(Ly$\alpha$) is the Ly$\alpha$ flux density in 10$^{-16}$ erg s$^{-1}$ cm$^{-2}$ measured in
the COS spectrum and corrected for the Milky Way extinction.}

\hbox{$^{\rm c}$$L$(Ly$\alpha$) is the Ly$\alpha$ luminosity in erg s$^{-1}$ corrected for the
Milky Way extinction.}

\hbox{$^{\rm d}$EW(Ly$\alpha$) is the rest-frame equivalent width in \AA\ of the Ly$\alpha$ emission line.}

\hbox{$^{\rm e}$$V_{\rm sep}$ is the Ly$\alpha$ velocity peak separation in km s$^{-1}$.}

\hbox{$^{\rm f}$$I$(H$\beta$) is the extinction-corrected H$\beta$ flux density in 
10$^{-16}$ erg s$^{-1}$ cm$^{-2}$ measured in the SDSS spectrum.}

\hbox{$^{\rm g}$$f_{\rm esc}$(Ly$\alpha$) is the ratio in percentage of 
$I$(Ly$\alpha$)/$I$(H$\beta$) to its case B value of 23.3.}

\hbox{$^{\rm h}$$f_{\rm esc}$(LyC) is the indirectly derived LyC escape fraction in
percentage, using the value of $V_{\rm sep}$ and equation 2 in \citet{I18b}.}

  \end{table*}

\begin{figure*}
\includegraphics[angle=0,width=0.90\linewidth]{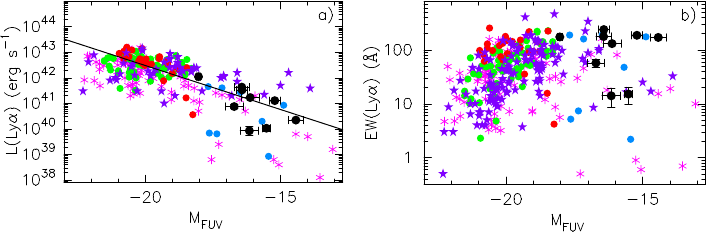}
\caption{Relations between the {\bf a)} Ly$\alpha$ luminosities 
$L$(Ly$\alpha$) and the absolute FUV magnitudes $M_{\rm FUV}$
derived from the rest-frame SEDs, {\bf b)} equivalent widths EW(Ly$\alpha$)
and the absolute FUV magnitudes $M_{\rm FUV}$. The relation in {\bf a)} is
defined by Eq.~\ref{amfuvllya}.
Meaning of symbols is the same as in Fig.~\ref{fig1}.
\label{fig6}}
\end{figure*}

\begin{figure*}
\includegraphics[angle=0,width=0.90\linewidth]{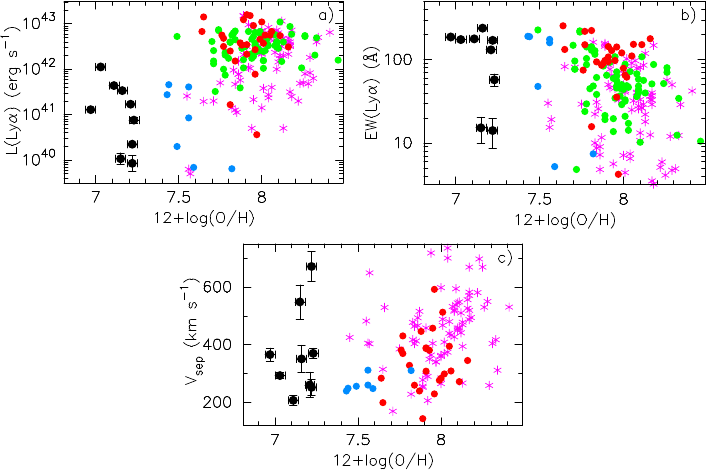}
\caption{Relations between the oxygen abundances and {\bf a)} Ly$\alpha$
luminosities $L$(Ly$\alpha$), {\bf b)} Ly$\alpha$ rest-frame equivalent
widths EW(Ly$\alpha$) and {\bf c)} Ly$\alpha$ peak separations
$V_{\rm sep}$. Meaning of symbols is the same as in Fig.~\ref{fig1}.
\label{fig7}}
\end{figure*}

\begin{figure*}
\includegraphics[angle=0,width=0.90\linewidth]{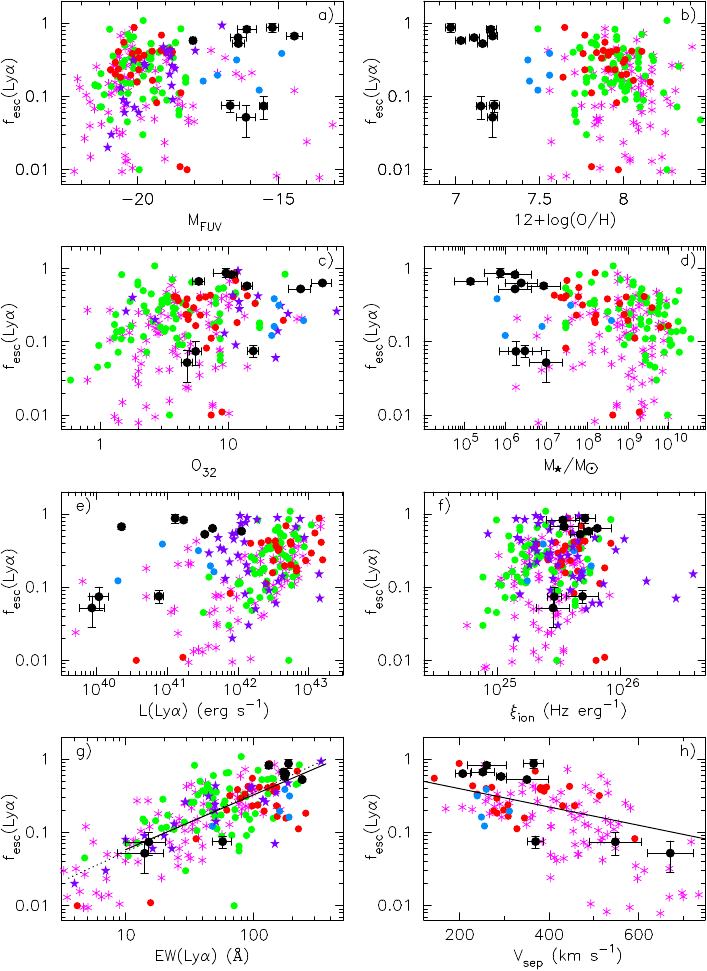}
\caption{Relations between the Ly$\alpha$ escape fractions 
$f_{\rm esc}$(Ly$\alpha$) and various integrated characteristics of the
low- and high-$z$ LAEs. Maximum likelihood relations in {\bf g)} and {\bf h)}
shown by solid lines are derived for the galaxies with
EW(Ly$\alpha$) $>$ 10\AA\ and $f_{\rm esc}$(Ly$\alpha$) $>$ 5 per cent and are
defined by Eqs.~\ref{fescEWlya} and \ref{fescVsep}, respectively.
On the other hand, the relation obtained for all galaxies is shown in
{\bf g)} with dotted line.
Meaning of symbols is the same as in Fig.~\ref{fig1}.
\label{fig8}}
\end{figure*}

\begin{figure*}
\includegraphics[angle=0,width=0.90\linewidth]{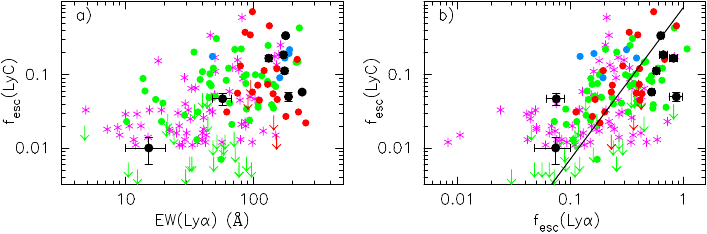}
\caption{Dependencies of the LyC escape fractions $f_{\rm esc}$(LyC) on {\bf a)}
equivalent widths of the Ly$\alpha$ emission line and {\bf b)} escape
fractions $f_{\rm esc}$(Ly$\alpha$) of the Ly$\alpha$ emission line. The
maximum likelihood relation in {\bf b)} is derived using only galaxies
with direct measurements of the Lyman continuum (red and green circles) and is
defined by Eq.~\ref{LyCLya}. Filled red and green circles
represent galaxies with detected Lyman continuum and arrows are
for the galaxies with upper limits,
respectively. The escape fractions of the Lyman continuum shown by blue, magenta
and black symbols
are derived using equation 2 by \citet{I18b} and excluding galaxies with
$V{\rm sep}$ $>$ 500 km s$^{-1}$.
\label{fig9}}
\end{figure*}

\section{Discussion}\label{discussion}

\subsection{Dependencies of the Ly$\alpha$ escape
fractions on Ly$\alpha$ equivalent widths and peak separations}

In this section we consider the properties derived from the Ly$\alpha$
emission of our xLAEs and compare them with the respective properties
of the low-$z$ and high-$z$ galaxies from the comparison sample. 
This sample includes the low-$z$ LyC leaking galaxies from \citet{I16,I16b,I18,I18b,I21a,I22},
\citet{Fl22}, the low-$z$ LAEs from \citet{I20}, \citet{H15}, \citet{JO14},
\citet{J17}, \citet{Y16,Y17}, \citet{Xu22}, \citet{Hu23}, and the high-$z$ LAEs
from the papers with references given in Table~\ref{tab4}.
For all low-$z$ LyC galaxies and LAEs from the comparison sample, we collected
and/or derived the same set of characteristics as for xLAEs.
All these galaxies have SDSS spectra, allowing
to derive some optical characteristics.
The data for high-$z$
LAEs are scarcer, with the richest data for the Ly$\alpha$
luminosities and equivalent widths.

In Fig.~\ref{fig6}a we display the $L$(Ly$\alpha$) -- $M_{\rm FUV}$ relation,
which shows a decline of the Ly$\alpha$ luminosity toward the fainter galaxies,
in line with the similar trend found by \citet{Ha17}.
The maximum likelihood fit to all data in the Figure, excluding
high-$z$ LAEs (purple stars), can be expressed as
\begin{equation}
\log L({\rm Ly}\alpha) = -0.34453 \times M_{\rm FUV} +
35.59659. \label{amfuvllya}
\end{equation}
The xLAEs (black symbols) extend the relation to fainter magnitudes.
We also note that faint $z$ $\ga$ 6 LAEs, with $M_{\rm FUV}$ $\ge$ --18 mag
(purple stars), are systematically brighter compared to low-$z$ LAEs.
The decline of Ly$\alpha$ luminosity is expected toward fainter small objects.
However, it is not as steep as the expected relation with a slope of $-$0.4 if
$L$(Ly$\alpha$) is just proportional to the UV luminosity. This is because
the equivalent widths EW(Ly$\alpha$) are higher for faint low-mass galaxies
(Fig.~\ref{fig6}b). The trend of increasing EW(Ly$\alpha$) with decreasing
$M_{\rm FUV}$ is also found for bright $z$ $\ga$ 6 galaxies with $M_{\rm FUV}$
$<$ --19 mag \citep{Jo23,Sa23}. An extension of the relation in Fig.~\ref{fig6}b
to fainter magnitudes, including xLAEs, reveals that the Ly$\alpha$ equivalent
widths at faint magnitudes $M_{\rm FUV}$ $\geq$ --18 mag become constant, with
the highest values not exceeding $\sim$ 250\AA. This is consistent with
the maximum Ly$\alpha$ equivalent widths of $\sim$ 240 -- 350\AA\ predicted
by standard synthesis models for metallicities between solar ($Z$ = 0.02) and
$Z$ = 0.0004 \citep{S03}.

The main goal of this study is to derive the properties of the Ly$\alpha$
emission in galaxies with the lowest known metallicities, and to search for possible
dependencies of the derived galaxy characteristics on metallicity. Such characteristics can be the Ly$\alpha$ luminosity,
the equivalent width EW(Ly$\alpha$), or the separation $V_{\rm sep}$ between the Ly$\alpha$
peaks. In Fig.~\ref{fig7}a, we show the relation
between oxygen abundance 12+log(O/H) and Ly$\alpha$ luminosity
$L$(Ly$\alpha$). A clear trend of increasing $L$(Ly$\alpha$) with increasing
12+log(O/H) is seen. This trend is somewhat expected because xLAEs being the
lowest-mass (Fig.~\ref{fig1}c) and faintest galaxies have also the lowest Ly$\alpha$
luminosities (Fig.~\ref{fig6}a). On the other hand, there are no evident
correlations of EW(Ly$\alpha$) and $V_{\rm sep}$ with 12+log(O/H)
(Fig.~\ref{fig7}b -- \ref{fig7}c). Thus, low metallicity is
likely not a good indicator for the selection of galaxies with high LyC leakage.
Seven xLAEs are characterised by high EW(Ly$\alpha$) (Fig.~\ref{fig7}b) and
small $V_{\rm sep}$ (Fig.~\ref{fig7}c), classifying them as possible LyC leakers.
On the other hand, two xLAEs have low EW(Ly$\alpha$) and high $V_{\rm sep}$,
indicating the presence of optically thick neutral gas in the line of sight between
the source of ionizing radiation and the observer.

In Fig.~\ref{fig8} are shown various dependencies of the Ly$\alpha$ escape
fraction on diverse characteristics of xLAEs and galaxies from the
comparison sample. The escape fraction $f_{\rm esc}$(Ly$\alpha$) is defined as the ratio of the
extinction-corrected Ly$\alpha$ to H$\beta$ flux ratio to its Case B value
of 23.3, corresponding to a low electron number density, 
$N_{\rm e}$ $\sim$ 10$^2$ cm$^{-3}$ \citep{SH95}:
\begin{equation}
 f_{\rm esc}({\rm Ly}\alpha) = 
\frac{1}{23.3}\frac{I({\rm Ly}\alpha)}{I({\rm H}\beta)}, \label{fesc}
\end{equation}
where $I$(Ly$\alpha$) is the flux corrected for the Milky Way 
extinction, and $I$(H$\beta$) is the flux
corrected for both the Milky Way and internal galaxy extinction.

Eq.~\ref{fesc} was applied to derive $f_{\rm esc}$(Ly$\alpha$) for both
the xLAE and comparison galaxies. Using both sets of data, we search for
possible correlations of $f_{\rm esc}$(Ly$\alpha$) with galaxy integrated
characteristics. It is seen in Fig.~\ref{fig8} that low-$z$ and high-$z$ LAEs
from the comparison sample overlap well, indicating that LAEs at reionization
epoch have properties very similar to those of nearby LAEs. An exception
is the xLAE sample. Galaxies from this sample are fainter and less massive, with lower
metallicities and Ly$\alpha$ luminosities. Other
characteristics, such as O$_{32}$ ratios and ionizing photon production
efficiencies $\xi_{\rm ion}$, are similar to those of the
galaxies in the comparison sample.

We find no or weak correlation between $f_{\rm esc}$(Ly$\alpha$) and absolute FUV
magnitude $M_{\rm FUV}$, oxygen abundance 12 + log(O/H), O$_{32}$ ratio, stellar
mass $M_\star$, Ly$\alpha$ luminosity and ionizing photon production
efficiency. On the other hand, relatively tight correlations are found
between $f_{\rm esc}$(Ly$\alpha$) and EW(Ly$\alpha$) (Fig.~\ref{fig8}g), and
between $f_{\rm esc}$(Ly$\alpha$) and 
$V_{\rm sep}$ (Fig.~\ref{fig8}h), in accord with the study by \citet{Y17}.
The maximum likelihood relation shown in Fig.~\ref{fig8}g by the dotted line
is obtained by including all galaxies. However, the errors of EW(Ly$\alpha$)
and $f_{\rm esc}$(Ly$\alpha$) might be very large for galaxies with weak
Ly$\alpha$ emission, due to their higher flux uncertainties, as indicated by the 
larger errors of our xLAEs (black symbols) at low EW(Ly$\alpha$). To study the
role of these uncertainties on the determination of the maximum likelihood
relation, we show in Fig.~\ref{fig8}g by a solid line the relation for
galaxies with EW(Ly$\alpha$)~$>$~10\AA\ and $f_{\rm esc}$(Ly$\alpha$)~$>$~5 per
cent. This relation expressed as
\begin{equation}
\log f_{\rm esc}({\rm Ly}\alpha) = 0.74851\times \log {\rm EW}({\rm Ly}\alpha) -
1.98411 \label{fescEWlya}
\end{equation}
is nearly coincident with the one for the entire sample. According to
Eq.~\ref{fescEWlya}, the ``intrinsic'' Ly$\alpha$ equivalent width
defined as EW(Ly$\alpha$)/$f_{\rm esc}$(Ly$\alpha$), i.e. the value of
EW(Ly$\alpha$) corresponding to the Ly$\alpha$ escape fraction of 1,
is equal to $\sim$~380\AA\ in galaxies
with the high observed EW(Ly$\alpha$) = 250\AA. This ``intrinsic'' value
is consistent with the values predicted by the population synthesis models
\citep[e.g. ][]{S03}.

The respective correlation between the Ly$\alpha$ peak separations and
Ly$\alpha$ escape fractions in Fig.~\ref{fig8}h for the restricted sample with
EW(Ly$\alpha$) $>$ 10\AA\ and $f_{\rm esc}$(Ly$\alpha$) $>$ 5 per cent
can be fit by the maximum likelihood equation in the form:
\begin{equation}
\log f_{\rm esc}({\rm Ly}\alpha) = -0.00125\times V_{\rm sep} -
0.15279. \label{fescVsep}
\end{equation}

\subsection{Correlation between the escape fractions of LyC and Ly$\alpha$ emission}

It is important for studies of the reoinization of the Universe to find
correlations between the escape fraction of the LyC emission $f_{\rm esc}$(LyC)
and some characteristics of the Ly$\alpha$ emission line.
One of the most commonly measured characteristics is the equivalent width
EW(Ly$\alpha$) of the Ly$\alpha$ emission line. As for $f_{\rm esc}$(LyC), it
can be obtained from the observed flux of the LyC emission in the COS spectra
of the galaxies with $z$ $\sim$ 0.3 -- 0.4. Concerning galaxies with
$z$~$<$~0.3, for which direct measurements of the LyC fluxes
in the {\sl HST}/COS spectra are not possible, the relation between the
$f_{\rm esc}$(LyC) and the Ly$\alpha$ peak separation can be used  \citep[e.g. ][]{I18b}.

In Fig.~\ref{fig9}a, we show the relation between $f_{\rm esc}$(LyC) and
EW(Ly$\alpha$) for the low-$z$ LyC leakers at $z$ $\sim$ 0.3 -- 0.4 and LAEs
at $z$ $\la$ 0.3, including our xLAEs. It is seen that the correlation between
$f_{\rm esc}$(LyC) and EW(Ly$\alpha$) is weak. We conclude 
that EW(Ly$\alpha$) is a poor indicator of LyC escaping emission.
The Figure
allows also to conclude that LAEs with EW(Ly$\alpha$) $\la$ 20\AA\ are mainly
non-leakers. 

A tighter correlation is found between $f_{\rm esc}$(LyC) and
$f_{\rm esc}$(Ly$\alpha$) (Fig.~\ref{fig9}b), where $f_{\rm esc}$(Ly$\alpha$) is
derived from the extinction-corrected flux ratios of the Ly$\alpha$ and H$\beta$
emission lines (Eq.~\ref{fesc}). In this Figure, direct observations
of the Lyman continuum are available for galaxies shown by red and green
filled circles (LyC is detected) and arrows (upper limits of the LyC
fluxes). For the remaining galaxies shown by blue, magenta and black symbols,
the LyC escape fraction is derived indirectly with the use of equation 2
by \citet{I18b}. We note however that the peak separations for all galaxies
but one, used in the determination of this relation, are less than
500 km s$^{-1}$. Thus, the indirectly derived LyC escape fraction in galaxies
with $V_{\rm sep}$ $>$ 500 km s$^{-1}$ may be badly determined and we excluded
these galaxies in Fig.~\ref{fig9}b. The maximum likelihood relation in
this Figure (solid line) is derived only for galaxies with direct measurements
of LyC (red and green filled and open circles). This relation takes the form:
\begin{equation}
\log f_{\rm esc}({\rm LyC}) = 2.06473 \times \log f_{\rm esc}({\rm Ly}\alpha) - 0.08773.
      \label{LyCLya}
\end{equation}
The distribution of galaxies with indirect determinations of
$f_{\rm esc}$(LyC) is similar to the distribution of galaxies with
direct measurements of the Lyman continuum fluxes. It follows from Eq.~\ref{LyCLya} that $f_{\rm esc}$(LyC) is always lower than $f_{\rm esc}$(Ly$\alpha$),
in agreement with conclusions by \citet{V17} and \citet{I21a}, theoretical
predictions by \citet*{D16} and numerical simulations by \citet{Ma22}.
The relation Eq.~\ref{LyCLya} can be used to determine $f_{\rm esc}$(LyC) in the
cases when direct measurements of the LyC fluxes are not possible
and resolved Ly$\alpha$ profiles, needed for the determination of
$V_{\rm sep}$, are not available. We note, however,
that this relation may not be valid if the opacity of the intergalactic medium in
the direction of the galaxy is high.

\section{Conclusions}\label{summary}

We present here {\sl HST}/COS observations of nine 
low-redshift ($z$ $\la$ 0.13) most metal-deficient compact star-forming
galaxies (SFG), with oxygen abundances 12$+$log(O/H) in the range 6.97~-- ~.23.
All studied objects are compact and low-mass
(log~($M_\star$/M$_\odot$)~=~5.2~--~7.0). They show strong nebular emission lines in
their optical spectra (EW(H$\beta$)~=~134~--~580\AA), indicating very 
young starburst ages of $<$~4~Myr.
We use these data to study the Ly$\alpha$ emission and indirect indicators
of the escaping Lyman continuum (LyC) radiation of these SFGs.
We compare their properties with those of the low-$z$
LyC galaxies and LAEs and high-$z$ ($z$ $>$ 6) LAEs, gathered from the literature.
Our main results are summarized as follows:

1. A Ly$\alpha$ emission line with two peaks was observed in the
spectra of all galaxies, classifying them as LAEs. 
This line is strong in seven galaxies, with high equivalent widths EW(Ly$\alpha$) $\ga$ 60\AA.
Our extremely low-metallicity LAEs (xLAEs) considerably extend the range of
physical properties of LAEs to
a faint UV magnitude of $-$14 mag, a low stellar mass of 10$^5$~M$_\odot$ and
a low metallicity 12+log(O/H)~$\sim$~7.0,  the lowest known so far.

2. We discuss various indirect indicators of escaping Ly$\alpha$ and
ionizing radiation. We find correlations between the Ly$\alpha$ escape fractions
and the equivalent widths EW(Ly$\alpha$) and the velocity separations $V_{\rm sep}$ between
the two peaks of the Ly$\alpha$ profile. We conclude that
both EW(Ly$\alpha$) and $V_{\rm sep}$ can be used as secondary indicators to determine
$f_{\rm esc}$(Ly$\alpha$). However, there is no correlation between oxygen
abundances and Ly$\alpha$ escape fractions.

3. We also find a correlation between the Ly$\alpha$ and LyC escape
fractions, which allows to estimate the LyC escape fractions when
direct observations of the Lyman continuum are not possible with the
{\sl HST}, e.g. in low-$z$ LAEs with $z$~$\leq$~0.3.

4. All our xLAEs are very compact
in the COS near ultraviolet (NUV) acquisition images. The surface brightness 
profiles at the outskirts of these galaxies can be approximated by an
exponential disc profile, 
with a scale length of $\sim$ 0.04 -- 0.21 kpc. These scale lengths are several 
times lower than those of confirmed LyC leakers and are among the lowest ones 
found for local blue compact dwarf galaxies. 

5. We find that the global properties of low-$z$ LAEs are very similar
to those of $z$ $>$ 6 galaxies. They are thus ideal nearby 
laboratories for investigating the mechanisms responsible for the
escape of Ly$\alpha$ and ionizing radiation from galaxies during the epoch
of the reionization of the Universe.

\section*{Acknowledgements}

These results are based on observations made with the NASA/ESA 
{\sl Hubble Space Telescope}, 
obtained from the data archive at the Space Telescope Science Institute. 
STScI is operated by the Association of Universities for Research in Astronomy,
Inc. under National Aeronautics and Space Administration (NASA) contract NAS 5-26555. Support for T.X.T. was provided by 
NASA through grant number HST-GO-16672.002-A from the Space Telescope Science 
Institute.
Y.I.I. and N.G.G. acknowledge support from the
National Academy of Sciences of Ukraine (Project No. 0123U102248) and from
the Simons Foundation. 
{\sc iraf} is distributed by 
the National Optical Astronomy Observatories, which are operated by the 
Association of Universities for Research in Astronomy, Inc., under cooperative 
agreement with the National Science Foundation.
{\sc stsdas} is a product of 
the Space Telescope Science Institute, which is operated by AURA for NASA.
Funding for the Sloan Digital Sky Survey IV has been provided by the 
Alfred P. Sloan Foundation, the U.S. Department of Energy Office of Science, 
and the Participating Institutions. SDSS-IV acknowledges
support and resources from the Center for High-Performance Computing at
the University of Utah. The SDSS web site is www.sdss.org.
SDSS-IV is managed by the Astrophysical Research Consortium for the 
Participating Institutions of the SDSS Collaboration.
{\sl Galaxy Evolution Explorer} ({\sl GALEX}) is a NASA mission  managed  by  the  Jet  Propulsion  Laboratory.
This research has made use of the NASA/Infrared Processing and Analysis Center
(IPAC) Extragalactic Database (NED) which 
is operated by the Jet  Propulsion  Laboratory,  California  Institute  of  
Technology,  under  contract with the National Aeronautics and Space 
Administration. This publication makes use of data products from the
{\sl Wide-field Infrared Survey Explorer} ({\sl WISE}), which is a joint project of the
University of California, Los Angeles, and the Jet Propulsion Laboratory,
California Institute of Technology, funded by the National Aeronautics and
Space Administration.






\section*{Data availability}

The data underlying this article will be shared on reasonable request to the 
corresponding author.

\appendix

\section{Emission-line fluxes in SDSS spectra and element abundances}

  \begin{table*}
  \caption{Extinction-corrected emission-line fluxes in SDSS spectra
\label{taba1}}
\begin{tabular}{lcrrrrr} \hline
 & &\multicolumn{5}{c}{100$\times$$I$($\lambda$)/$I$(H$\beta$)$^{\rm a}$}\\
Line &\multicolumn{1}{c}{$\lambda$}&J0122$+$0048&J0139$+$1542&J0811$+$4730&J0837$+$1921&J1004$+$3256\\
\hline
$[$O~{\sc ii}$]$     &3727&  51.6$\pm$~\,5.5& 50.3$\pm$~\,4.7& 17.3$\pm$~\,3.0& 33.1$\pm$~\,4.9&  9.0$\pm$~\,1.9\\
H12                  &3750&       ...       &      ...       &       ...      &      ...       &      ...       \\
H11                  &3771&       ...       &      ...       &       ...      &      ...       &  5.3$\pm$~\,1.4\\
H10                  &3798&       ...       &      ...       &  6.9$\pm$~\,2.7&      ...       &  8.0$\pm$~\,1.6\\
H9                   &3836&       ...       &      ...       &  8.2$\pm$~\,2.6&      ...       &  8.2$\pm$~\,1.6\\
$[$Ne~{\sc iii}$]$   &3869&  22.6$\pm$~\,4.0& 30.6$\pm$~\,3.9& 14.7$\pm$~\,2.9& 36.9$\pm$~\,4.8& 31.2$\pm$~\,2.7\\
H8+He~{\sc i}        &3889&  25.3$\pm$~\,4.1& 21.9$\pm$~\,3.3& 22.0$\pm$~\,3.1& 20.6$\pm$~\,3.9& 20.5$\pm$~\,2.2\\
H7+$[$Ne~{\sc iii}$]$&3969&  26.9$\pm$~\,4.1& 26.3$\pm$~\,3.6& 32.4$\pm$~\,3.7& 20.0$\pm$~\,3.8& 28.0$\pm$~\,2.5\\
H$\delta$            &4101&  29.0$\pm$~\,4.1& 29.2$\pm$~\,3.4& 27.5$\pm$~\,3.3& 31.5$\pm$~\,4.4& 29.2$\pm$~\,2.5\\
H$\gamma$            &4340&  49.2$\pm$~\,4.9& 50.4$\pm$~\,4.4& 49.3$\pm$~\,4.1& 51.9$\pm$~\,5.2& 47.5$\pm$~\,3.1\\
$[$O~{\sc iii}$]$    &4363&   9.4$\pm$~\,3.0& 12.9$\pm$~\,2.6&  6.5$\pm$~\,2.0& 16.8$\pm$~\,3.4& 15.2$\pm$~\,1.8\\
He~{\sc i}           &4471&       ...       &  4.1$\pm$~\,2.2&  3.6$\pm$~\,1.6&      ...       &  3.7$\pm$~\,1.1\\
He~{\sc ii}          &4686&       ...       &      ...       &  2.3$\pm$~\,1.1&  7.7$\pm$~\,2.8&  1.7$\pm$~\,1.0\\
H$\beta$             &4861& 100.0$\pm$~\,6.6&100.0$\pm$~\,5.8&100.0$\pm$~\,5.6&100.0$\pm$~\,6.9&100.0$\pm$~\,4.4\\
$[$O~{\sc iii}$]$    &4959&  86.1$\pm$~\,6.1&100.0$\pm$~\,5.8& 57.9$\pm$~\,4.2&116.0$\pm$~\,7.3&116.1$\pm$~\,4.8\\
$[$O~{\sc iii}$]$    &5007& 247.5$\pm$10.8  &297.9$\pm$11.0  &166.0$\pm$~\,7.4&353.0$\pm$14.1  &332.7$\pm$~\,9.2\\
He~{\sc i}           &5876&  10.7$\pm$~\,2.5& 15.0$\pm$~\,2.1&  8.7$\pm$~\,1.7&  9.8$\pm$~\,2.3& 11.3$\pm$~\,1.4\\
$[$O~{\sc i}$]$      &6300&   1.3$\pm$~\,1.6&      ...       &  1.3$\pm$~\,0.7&      ...       &      ...       \\
$[$S~{\sc iii}$]$    &6312&   0.6$\pm$~\,1.4&      ...       &      ...       &      ...       &      ...       \\
H$\alpha$            &6563& 274.1$\pm$12.1  &275.6$\pm$10.7  &275.3$\pm$10.6  &270.4$\pm$12.4  &149.1$\pm$~\,5.8$^{\rm b}$\\
$[$N~{\sc ii}$]$     &6583&   2.7$\pm$~\,1.6&  3.0$\pm$~\,1.1&  0.6$\pm$~\,0.7&  1.8$\pm$~\,1.6&  0.4$\pm$~\,0.3\\
He~{\sc i}           &6678&      ...        &  3.1$\pm$~\,1.2&  2.4$\pm$~\,1.0&      ...       &  2.5$\pm$~\,0.7\\
$[$S~{\sc ii}$]$     &6717&   3.8$\pm$~\,1.7&  6.1$\pm$~\,1.5&  1.8$\pm$~\,1.0&      ...       &  1.2$\pm$~\,0.6\\
$[$S~{\sc ii}$]$     &6731&   3.5$\pm$~\,1.7&  3.1$\pm$~\,1.2&  1.4$\pm$~\,0.9&      ...       &  1.2$\pm$~\,0.6\\
He~{\sc i}           &7065&       ...       &  5.4$\pm$~\,1.3&  3.2$\pm$~\,1.0&      ...       &  6.1$\pm$~\,1.0\\
$[$Ar~{\sc iii}$]$   &7136&       ...       &  2.0$\pm$~\,0.4&      ...       &      ...       &      ...       \\
$C$(H$\beta$)$^{\rm c}$   &&\multicolumn{1}{c}{0.080$\pm$0.053}&\multicolumn{1}{c}{0.195$\pm$0.047}&\multicolumn{1}{c}{0.135$\pm$0.045}&\multicolumn{1}{c}{0.070$\pm$0.054}&\multicolumn{1}{c}{0.180$\pm$0.047}\\
EW(H$\beta$)$^{\rm d}$      &&\multicolumn{1}{c}{156$\pm$10}&\multicolumn{1}{c}{339$\pm$34}&\multicolumn{1}{c}{331$\pm$30}&\multicolumn{1}{c}{134$\pm$13}&\multicolumn{1}{c}{459$\pm$46}\\
EW([O~{\sc iii}]$\lambda$5007)$^{\rm d}$ &&\multicolumn{1}{c}{496$\pm$19}&\multicolumn{1}{c}{993$\pm$99}&\multicolumn{1}{c}{590$\pm$43}&\multicolumn{1}{c}{726$\pm$73}&\multicolumn{1}{c}{1019$\pm$102}\\
EW(H$\alpha$)$^{\rm d}$     &&\multicolumn{1}{c}{1073$\pm$41}&\multicolumn{1}{c}{1950$\pm$195}&\multicolumn{1}{c}{1787$\pm$52}&\multicolumn{1}{c}{976$\pm$98}&\multicolumn{1}{c}{1105$\pm$111}\\
$F$(H$\beta$)$^{\rm e}$     &&\multicolumn{1}{c}{7.2$\pm$0.4}&\multicolumn{1}{c}{5.0$\pm$0.3}&\multicolumn{1}{c}{9.8$\pm$1.3}&\multicolumn{1}{c}{6.7$\pm$0.5}&\multicolumn{1}{c}{16.7$\pm$0.7}\\
\hline
  \end{tabular}

\hbox{$^{\rm a}$$I$($\lambda$) and $I$(H$\beta$) are emission-line
fluxes, corrected for the extinction derived from the Balmer decrement of hydrogen lines.}

\hbox{$^{\rm b}$Clipped line.}

\hbox{$^{\rm c}$$C$(H$\beta$) is the extinction coefficient derived from the Balmer decrement of
hydrogen lines.}



\hbox{$^{\rm d}$Rest-frame equivalent width in \AA.}

\hbox{$^{\rm e}$$F$(H$\beta$) is the observed H$\beta$ flux density in 10$^{-16}$ erg s$^{-1}$ cm$^{-2}$.}

  \end{table*}

\setcounter{table}{0}
\renewcommand{\thetable}{A\arabic{table}}

  \begin{table*}
  \caption{{\it continued} Extinction-corrected emission-line fluxes in SDSS spectra
}
\begin{tabular}{lcrrrr} \hline
 & &\multicolumn{4}{c}{100$\times$$I$($\lambda$)/$I$(H$\beta$)$^{\rm a}$}\\
Line &\multicolumn{1}{c}{$\lambda$}&J1206$+$5007&J1234$+$3901&J1505$+$3721&J2229$+$2725\\
\hline
$[$O~{\sc ii}$]$     &3727&  44.4$\pm$~\,5.0& 13.1$\pm$~\,2.4& 21.3$\pm$~\,2.0&  5.7$\pm$~\,1.3\\
H12                  &3750&       ...       &      ...       &  4.2$\pm$~\,1.4&  4.1$\pm$~\,1.4\\
H11                  &3771&       ...       &      ...       &  6.4$\pm$~\,1.4&  4.8$\pm$~\,1.4\\
H10                  &3798&       ...       &  6.9$\pm$~\,2.0&  7.5$\pm$~\,1.4&  6.0$\pm$~\,1.4\\
H9                   &3836&       ...       &  7.5$\pm$~\,1.9& 11.4$\pm$~\,1.6&  6.9$\pm$~\,1.5\\
$[$Ne~{\sc iii}$]$   &3869&  23.4$\pm$~\,3.9& 17.4$\pm$~\,2.5& 32.9$\pm$~\,2.3& 22.4$\pm$~\,2.1\\
H8+He~{\sc i}        &3889&  23.8$\pm$~\,3.8& 22.6$\pm$~\,2.7& 24.8$\pm$~\,2.0& 19.5$\pm$~\,2.0\\
H7+$[$Ne~{\sc iii}$]$&3969&  17.6$\pm$~\,3.6& 23.4$\pm$~\,2.8& 29.8$\pm$~\,2.1& 23.4$\pm$~\,2.1\\
H$\delta$            &4101&  27.2$\pm$~\,4.0& 27.7$\pm$~\,2.9& 33.3$\pm$~\,2.2& 26.0$\pm$~\,2.2\\
H$\gamma$            &4340&  46.0$\pm$~\,4.6& 47.7$\pm$~\,3.6& 51.2$\pm$~\,2.7& 46.3$\pm$~\,2.9\\
$[$O~{\sc iii}$]$    &4363&  11.1$\pm$~\,3.0&  6.9$\pm$~\,1.8& 14.7$\pm$~\,1.5& 14.0$\pm$~\,1.6\\
He~{\sc i}           &4471&       ...       &  3.3$\pm$~\,1.4&  2.9$\pm$~\,1.0&  3.9$\pm$~\,1.1\\
He~{\sc ii}          &4686&       ...       &  3.1$\pm$~\,1.4&  4.7$\pm$~\,1.1&  2.1$\pm$~\,0.8\\
H$\beta$             &4861& 100.0$\pm$~\,6.5&100.0$\pm$~\,5.1&100.0$\pm$~\,3.9&100.0$\pm$~\,4.3\\
$[$O~{\sc iii}$]$    &4959&  82.9$\pm$~\,5.9& 64.5$\pm$~\,4.0&116.8$\pm$~\,4.2&100.3$\pm$~\,4.3\\
$[$O~{\sc iii}$]$    &5007& 250.8$\pm$10.9  &185.0$\pm$~\,7.0&335.6$\pm$~\,4.7&312.3$\pm$~\,9.2\\
He~{\sc i}           &5876&   9.7$\pm$~\,2.3& 10.0$\pm$~\,1.6&  9.2$\pm$~\,1.1&  9.5$\pm$~\,1.3\\
$[$O~{\sc i}$]$      &6300&       ...       &      ...       &      ...       &  0.7$\pm$~\,0.5\\
$[$S~{\sc iii}$]$    &6312&       ...       &      ...       &      ...       &      ...       \\
H$\alpha$            &6563& 273.1$\pm$11.8  &246.7$\pm$~\,9.1&272.7$\pm$~\,8.0&270.9$\pm$~\,8.8\\
$[$N~{\sc ii}$]$     &6583&  1.6$\pm$~\,0.7 &  0.7$\pm$~\,0.4&  1.1$\pm$~\,0.6&  0.3$\pm$~\,0.2\\
He~{\sc i}           &6678&      ...        &  3.7$\pm$~\,1.0&  2.4$\pm$~\,0.6&  2.3$\pm$~\,0.7\\
$[$S~{\sc ii}$]$     &6717&   3.2$\pm$~\,1.5&      ...       &      ...       &      ...       \\
$[$S~{\sc ii}$]$     &6731&   3.0$\pm$~\,1.5&      ...       &      ...       &      ...       \\
He~{\sc i}           &7065&       ...       &  4.5$\pm$~\,1.1&  4.7$\pm$~\,0.8&  4.5$\pm$~\,0.9\\
$[$Ar~{\sc iii}$]$   &7136&       ...       &      ...       &  1.6$\pm$~\,0.6&  0.8$\pm$~\,0.5\\
$C$(H$\beta$)$^{\rm c}$   &&\multicolumn{1}{c}{0.125$\pm$0.052}&\multicolumn{1}{c}{0.145$\pm$0.044}&\multicolumn{1}{c}{0.100$\pm$0.036}&\multicolumn{1}{c}{0.060$\pm$0.039}\\
EW(H$\beta$)$^{\rm d}$      &&\multicolumn{1}{c}{219$\pm$15}&\multicolumn{1}{c}{276$\pm$15}&\multicolumn{1}{c}{298$\pm$30}&\multicolumn{1}{c}{580$\pm$48}\\
EW([O~{\sc iii}]$\lambda$5007)$^{\rm d}$ &&\multicolumn{1}{c}{429$\pm$18}&\multicolumn{1}{c}{610$\pm$24}&\multicolumn{1}{c}{351$\pm$35}&\multicolumn{1}{c}{1884$\pm$24}\\
EW(H$\alpha$)$^{\rm d}$     &&\multicolumn{1}{c}{1144$\pm$50}&\multicolumn{1}{c}{522$\pm$19}&\multicolumn{1}{c}{1717$\pm$172}&\multicolumn{1}{c}{2543$\pm$206}\\
$F$(H$\beta$)$^{\rm e}$     &&\multicolumn{1}{c}{7.4$\pm$0.5}&\multicolumn{1}{c}{12.4$\pm$0.7}&\multicolumn{1}{c}{24.7$\pm$0.9}&\multicolumn{1}{c}{17.8$\pm$0.7}\\
\hline
  \end{tabular}

\hbox{$^{\rm a}$$I$($\lambda$) and $I$(H$\beta$) are emission-line
fluxes, corrected for the extinction derived from the Balmer decrement of hydrogen lines.}

\hbox{$^{\rm b}$Clipped line.}

\hbox{$^{\rm c}$$C$(H$\beta$) is the extinction coefficient derived from the Balmer decrement of
hydrogen lines.}



\hbox{$^{\rm d}$Rest-frame equivalent width in \AA.}

\hbox{$^{\rm e}$$F$(H$\beta$) is the observed H$\beta$ flux density in 10$^{-16}$ erg s$^{-1}$ cm$^{-2}$.}

  \end{table*}

  \begin{table*}
  \caption{Electron temperatures, electron number densities and 
element abundances in H~{\sc ii} regions \label{taba2}}
  \begin{tabular}{lccccc} \hline
&\multicolumn{5}{c}{Galaxy} \\ 
Parameter                              &J0122$+$0048    &J0139$+$1542    &J0811$+$4730    &J0837$+$1921&J1004$+$3256 \\
\hline
$T_{\rm e}$ ($[$O {\sc iii}$]$), K        & 21540$\pm$4520 & 23510$\pm$3480 & 21900$\pm$4580 & 25000$\pm$4100& 24310$\pm$2200 \\
$T_{\rm e}$ ($[$O {\sc ii}$]$), K         & 15610$\pm$3000 & 15250$\pm$1950 & 15570$\pm$2940 & 14730$\pm$1900& 15000$\pm$1120 \\
$T_{\rm e}$ ($[$S {\sc iii}$]$), K        & 19570$\pm$3750 & 20430$\pm$2890 & 19820$\pm$3800 & 20870$\pm$3400& 20730$\pm$1820 \\
$N_{\rm e}$ ($[$S {\sc ii}$]$), cm$^{-3}$         &   500$\pm$500  &   10$\pm$10 &   110$\pm$110  &   678$\pm$678  \\ \\
O$^+$/H$^+$$\times$10$^{5}$            &0.44$\pm$0.08  &0.43$\pm$0.06  &0.14$\pm$0.03  &0.32$\pm$0.05&0.09$\pm$0.02  \\
O$^{2+}$/H$^+$$\times$10$^{5}$         &1.24$\pm$0.08  &1.24$\pm$0.08  &0.80$\pm$0.06  &1.31$\pm$0.11&1.32$\pm$0.06  \\
O$^{3+}$/H$^+$$\times$10$^{6}$         &     ...       &     ...       &     ...       &      ...    &0.29$\pm$0.18  \\
O/H$\times$10$^{5}$                   &1.68$\pm$0.12  &1.68$\pm$0.10  &0.94$\pm$0.07  &1.63$\pm$0.12&1.44$\pm$0.07 \\
12+log(O/H)                           &7.22$\pm$0.03  &7.22$\pm$0.03  &6.97$\pm$0.03  &7.21$\pm$0.03&7.16$\pm$0.02 \\ \\
N$^+$/H$^+$$\times$10$^{7}$            &      ...      &2.14$\pm$0.85  &0.42$\pm$0.05 &      ...      \\
ICF(N)$^{\rm a}$                       &      ...      &3.81           &6.25          &      ...       \\
N/H$\times$10$^{6}$                   &      ...      &0.81$\pm$0.33  &0.26$\pm$0.32 &      ...       \\
log(N/O)                              &      ...      &~$-$1.31$\pm$0.18~~\,&~$-$1.56$\pm$0.53~~\,&  ... \\ \\
Ne$^{2+}$/H$^+$$\times$10$^{5}$        &0.24$\pm$0.05  &0.27$\pm$0.03  &0.15$\pm$0.03  &0.29$\pm$0.04&0.26$\pm$0.02 \\
ICF(Ne)$^{\rm a}$                      &1.11           &1.11           &1.06           &1.08         &1.04           \\
Ne/H$\times$10$^{5}$                  &0.27$\pm$0.05  &0.30$\pm$0.04  &0.16$\pm$0.03  &0.31$\pm$0.04&0.27$\pm$0.02 \\
log(Ne/O)                            &~$-$0.79$\pm$0.09~~\,&~$-$0.74$\pm$0.06~~\,&~$-$0.77$\pm$0.09~~\,&~$-$0.71$\pm$0.07~~\,&~$-$0.73$\pm$0.04~~\,\\ \\
S$^{+}$/H$^+$$\times$10$^{7}$          &0.69$\pm$0.22  &     ...       &     ...       &     ... &     ...   \\
S$^{2+}$/H$^+$$\times$10$^{6}$         &0.15$\pm$0.36  &      ...      &     ...       &     ... &     ...    \\
ICF(S)$^{\rm a}$                       &1.16           &     ...       &     ...       &    ...  &     ...    \\
S/H$\times$10$^{6}$                   &0.25$\pm$0.42  &     ...       &     ...       &    ... &     ...  \\
log(S/O)                             &~$-$1.82$\pm$0.73~~\,&    ...  & ...          &      ...  &     ...    \\ \hline
&\multicolumn{4}{c}{Galaxy} \\ 
Parameter                              &J1206$+$5007    &J1234$+$3901    &J1505$+$3721    &J2229$+$2725\\ 
\cline{1-5}
$T_{\rm e}$ ($[$O {\sc iii}$]$), K        & 23890$\pm$4840 & 21210$\pm$3540 & 23570$\pm$1770 & 24260$\pm$2170 \\
$T_{\rm e}$ ($[$O {\sc ii}$]$), K         & 15140$\pm$2600 & 15630$\pm$2380 & 15240$\pm$980 & 15020$\pm$1120  \\
$T_{\rm e}$ ($[$S {\sc iii}$]$), K        & 20610$\pm$4010 & 19460$\pm$2930 & 20490$\pm$1470 & 20710$\pm$1800 \\
$N_{\rm e}$ ($[$S {\sc ii}$]$), cm$^{-3}$         &   471$\pm$471  &  10$\pm$10 &   10$\pm$10 & 10$\pm$10 \\ \\
O$^+$/H$^+$$\times$10$^{5}$            &0.41$\pm$0.07  &0.10$\pm$0.02  &0.19$\pm$0.02  &0.05$\pm$0.01 \\
O$^{2+}$/H$^+$$\times$10$^{5}$         &1.01$\pm$0.09  &0.95$\pm$0.06  &1.41$\pm$0.05  &1.22$\pm$0.06 \\
O$^{3+}$/H$^+$$\times$10$^{6}$         &      ...      &0.26$\pm$0.14  &0.92$\pm$0.33  &0.34$\pm$0.17 \\
O/H$\times$10$^{5}$                   &1.42$\pm$0.11  &1.08$\pm$0.06  &1.68$\pm$0.06  &1.30$\pm$0.06 \\
12+log(O/H)                           &7.15$\pm$0.04  &7.03$\pm$0.02  &7.23$\pm$0.02  &7.11$\pm$0.02 \\ \\
Ne$^{2+}$/H$^+$$\times$10$^{5}$        &0.20$\pm$0.03  &0.19$\pm$0.03  &0.29$\pm$0.02  &0.19$\pm$0.02 \\
ICF(Ne)$^{\rm a}$                      &1.12           &1.05           &1.07           &1.03  \\
Ne/H$\times$10$^{5}$                  &0.23$\pm$0.04  &0.20$\pm$0.03  &0.31$\pm$0.02  &0.19$\pm$0.02 \\
log(Ne/O)                             &~$-$0.80$\pm$0.08~~\,&~$-$0.73$\pm$0.07~~\,&~$-$0.73$\pm$0.03~~\,&~$-$0.82$\pm$0.05~~\, \\ \\
Ar$^{2+}$/H$^+$$\times$10$^{7}$        &     ...       &      ...      &0.39$\pm$0.13  &0.19$\pm$0.12 \\
ICF(Ar)$^{\rm a}$                      &     ...       &      ...      &1.33           &2.14 \\
Ar/H$\times$10$^{7}$                  &     ...       &      ...      &0.52$\pm$0.18  &0.41$\pm$0.69 \\
log(Ar/O)                            &     ...       &      ...      &~$-$2.51$\pm$0.15~~\,&~$-$2.50$\pm$0.73~~\, \\ 
\cline{1-5}
\end{tabular}

\hbox{$^{\rm a}$ICF is the ionization correction factor.}
  \end{table*}

\bsp	
\label{lastpage}
\end{document}